\documentclass[acmsmall]{acmart}
\usepackage{multirow}
\usepackage{graphicx}
\usepackage{makecell}
\usepackage{pifont}
\usepackage{enumitem}

\AtBeginDocument{%
  }

\setcopyright{acmlicensed}
\copyrightyear{2025}
\acmYear{2025}
\acmDOI{XXXXXXX.XXXXXXX}
\acmJournal{TOSEM}

\begin{document}

\title{Artificial Intelligence Support for Software Architecture Practice: A Systematic Review and Future Directions}

\author{Alessio Bucaioni}
\orcid{https://orcid.org/0000-0002-8027-0611}
\email{alessio.bucaioni@mdu.se}
\affiliation{%
  \institution{Mälardalen Universty}
  \country{Sweden}
}
\author{Martin Weyssow}
\orcid{https://orcid.org/0000-0002-5987-850X}
\email{mweyssow@smu.edu.sg}

\author{Junda He}
\orcid{https://orcid.org/0000-0003-3370-8585}
\email{jundahe@smu.edu.sg}
\author{Yunbo Lyu}
\orcid{https://orcid.org/0009-0004-2522-7348}
\email{yunbolyu@smu.edu.sg}
\author{David Lo}
\orcid{https://orcid.org/0000-0002-4367-7201}
\email{davidlo@smu.edu.sg}
\affiliation{%
  \institution{Singapore Management University}
  \country{Singapore}
}

\renewcommand{\shortauthors}{Bucaioni et al.}

\begin{abstract}
Artificial intelligence is increasingly applied across software engineering, yet its explicit role in software architecture remains insufficiently understood. Architectural practices rely on complex trade-offs, documentation, and long-term evolution, all of which are traditionally manual, error-prone, and difficult to sustain. 
To clarify how artificial intelligence can address these challenges, we conducted a systematic literature review of 51 peer-reviewed primary studies and systematically mapped their contributions onto 17 practitioner-reported software architecture challenges derived from empirical interviews. This analysis identifies 14 topical areas where artificial intelligence has been applied to architectural tasks and idetifies six artificial intelligence–specific challenges that expose fundamental gaps between current capabilities and practitioner needs. 
Building on these findings, we chart a research agenda for artificial intelligence–driven software architecture organized around five strategic pillars. 
By grounding the roadmap in both systematic evidence and practitioner insights, this work provides the first peer-reviewed comprehensive synthesis of artificial intelligence contributions to software architecture, establishes a foundation for future research, and outlines the conditions under which artificial intelligence can become a trustworthy partner in architectural design, evaluation, and evolution.

\end{abstract}



\received{February 2025}
\received[revised]{June 2026}

\maketitle

\section{Introduction}
The growing impact of Artificial Intelligence (AI) on Software Engineering (SE) has been widely documented~\cite{yang2024robustness}, particularly in tasks operating at the code level, such as code generation~\cite{dehaerne2022code, zhuo2024bigcodebench, DBLP:journals/corr/abs-2508-12285}, bug detection~\cite{pradel2018deepbugs, li2024enhancing}, and program repair~\cite{li2022dear, zhou2024leveraging}. In contrast, the explicit use of AI in Software Architecture (SA) remains sparse. This is surprising, given that SA forms the blueprint upon which robust, scalable, and secure systems are built~\cite{bass2012software}.

Despite its importance, architectural practice remains largely manual and experience-driven, demanding deep domain knowledge and sustained reasoning about long-term trade-offs~\cite{Bucaioni7118}. Architects must continually balance qualities such as performance, security, and reliability under changing constraints. Yet in many industrial contexts, architectural investment is undervalued in the face of short-term time-to-market pressures~\cite{xiao2016identifying}, even though neglecting SA inevitably compromises long-term system quality and maintainability~\cite{bass2012software}.

The emerging promise of AI presents an opportunity to address these challenges by reducing the cognitive burden on architects and automating aspects of architectural work. Early evidence suggests AI can derive partial design solutions from requirements~\cite{P2}, perform trade-off analyses~\cite{P20}, and even prioritize quality attributes~\cite{P26,P34}. However, compared to programming tasks, architectural decision-making requires abstraction, long-term reasoning, and the integration of diverse and often conflicting quality attributes—capabilities that current AI techniques struggle to fully support.

This gap motivates a central research question: \emph{How can AI be effectively leveraged to overcome the inherent difficulties of architectural tasks without oversimplifying the complexity that SA demands?} Answering this question is not only a technological challenge but also a strategic one. As Ozkaya notes~\cite{ozkaya2023next}, the move toward AI-enhanced architecture should be driven by a sober understanding of challenges rather than by a “fear of missing out.”

In this paper, we make the following contributions:
\begin{itemize}
    \item We conduct a systematic literature review (SLR) of AI applications in SA, identifying 14 state-of-the-art practices and techniques.
    \item We integrate these findings with 32 practitioner-elicited architectural challenges~\cite{wan2023software}, providing the first peer-reviewed systematic mapping of AI contributions against real-world needs.
    \item We identify six AI-specific challenges that remain unsolved, highlighting where current approaches fall short.
    \item We propose a forward-looking vision and roadmap for AI in SA (AI4SA), outlining five strategic research directions that can guide both researchers and practitioners.
\end{itemize}

While this paper represents, to the best of our knowledge, the first peer-reviewed roadmap specifically focused on the role of AI in software architecture and grounded in a systematic literature review and empirical findings, we fully recognize its complementarity with seminal roadmaps in the field~\cite{garlan2014software, kramer2007self}. 
Prior roadmap contributions have provided broad and influential research directions for software architecture, covering issues such as architectural evolution, self-adaptation, socio-technical ecosystems, education, and the changing role of architects~\cite{garlan2014software}. Our contribution is complementary: rather than proposing a general roadmap for the entire software architecture field, we focus specifically on how AI can support software architecture practice, including design, decision-making, documentation, analysis, maintenance, and evolution activities.

The remainder of the paper is structured as follows.  
Section~\ref{sec:met} presents the research methodology.  
Section~\ref{sec:synt} synthesizes the insights from our review.  
Section~\ref{sec:ch} evaluates the contributions of AI to software architecture and identifies the persistent challenges that remain. 
Section~\ref{sec:tools} we examine a set of mainstream tools used in contemporary architectural workflows and relate these capabilities to the identified persistent challenges.
Section~\ref{sec:roa} outlines a vision for AI4SA through five strategic pillars for future advancement.  
Section~\ref{sec:dis} discusses the main findings and their implications.  
Section~\ref{sec:threats} and Section~\ref{sec:related} address the threats to validity and review related work, respectively.  
Finally, Section~\ref{sec:conclusion} concludes the paper with final remarks and directions for future research.

\section{Research Process} \label{sec:met}
We built our forward-looking vision for AI4SA using a  two-phase research process, as follows.

First, we conducted a SLR of the current applications of AI in SA to map state-of-the-art practices, techniques, and reported outcomes. The SLR was designed and executed following the guidelines for conducting secondary studies in software engineering proposed by Kitchenham et al.~\cite{kitchenham2013systematic}. The SLR process involved three key phases: planning, conducting, and documenting. In the planning phase, we defined the Research Goal (RG), formulated the Research Questions (RQs), and established a research protocol to ensure methodological rigour and replicability. The conducting phase involved executing the protocol, including search strategy design, study selection, data extraction schema definition, data extraction, and data analysis. The documenting phase focused on recording results and analysing potential threats to validity to support transparency and reliability.

Second, we synthesized the findings of the SLR with a set of open SA challenges identified in our prior work~\cite{wan2023software}. This integration allowed us to assess the extent and nature of AI’s contributions to addressing these challenges. Based on this synthesis, we identified AI-specific research gaps and articulated a vision comprising six strategic directions for advancing AI4SA.

To facilitate independent replication and verification, we provide a complete replication package in Section~\ref{rep} containing search and selection data, and the list of primary studies.

\subsection{Research Goal}
Following the Goal-Question-Metric (GQM) approach~\cite{gqm}, we defined the Research Goal (RG) guiding our SLR, as presented in Table~\ref{tab:gqm}.

\begin{table*}[!htbp]
    \centering
    \caption{Research goal expressed using the GQM perspectives.}
    \label{tab:gqm}
    \begin{tabular}{ p{1.5cm} | p{9cm} }
    \textit{Purpose} & Identify, and classify  \\ 
    \textit{Issue} & Techniques and applications\\ 
    \textit{Object} & AI for SA\\ 
    \textit{Viewpoint} & From the point of view of researchers and practitioners. \\ 
    \end{tabular}
\end{table*}

This RG, structured through the GQM approach, scopes and guides our SLR by framing research questions that identify and classify the current landscape of AI techniques and applications in software architecture. Specifically, the issue dimension covers both methodological approaches and applied use cases, the object of study comprises AI-related practices, tools, and frameworks supporting architectural activities, and the viewpoint reflects both academic and industrial perspectives to capture a holistic understanding of the field. Accordingly, while earlier self-adaptive architecture research primarily addressed runtime adaptation under uncertainty, our review focuses on the growing impact of contemporary AI techniques and on how they support architectural tasks across the lifecycle, beyond runtime control mechanisms.

\subsection{Search and selection}
\begin{figure}[!h]
\centering
    \includegraphics[width=\linewidth]{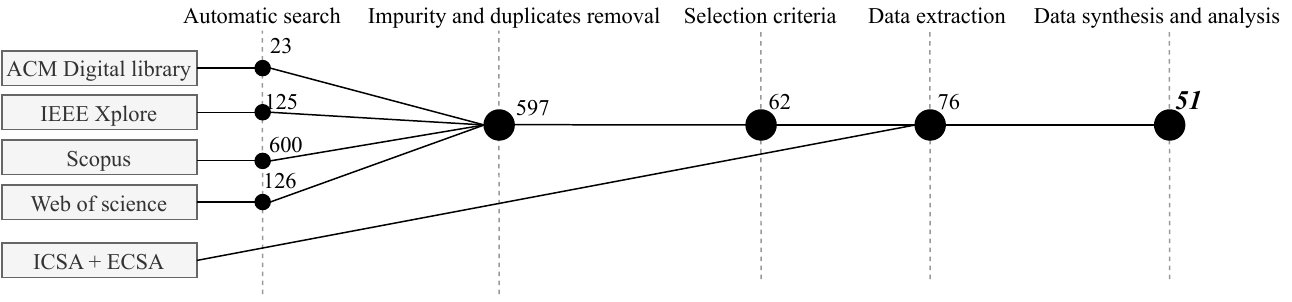}
    \caption{Search and selection}
    \label{fig:ss}
\end{figure}

Following the steps illustrated in Figure~\ref{fig:ss}, we collected relevant research studies for our investigation. We started with an automatic search of four of the largest scientific databases and indexing systems in software engineering~\cite{kitchenham2013systematic}: \textit{IEEE Xplore Digital Library}, \textit{ACM Digital Library}, \textit{SCOPUS} and \textit{Web of Science} (Table~\ref{tab:db}).

\begin{table}[]
\caption{Electronic databases and indexing systems used in this study}
\label{tab:db}
\begin{center}
   \begin{tabular}{| l | l | l |} 
   \hline
    \textbf{Name} & \textbf{Type} & \textbf{URL} \\ \hline   \hline
   IEEE Xplore Digital Library & Electronic database &  \url{http://ieeexplore.ieee.org} \\ \hline  
   ACM Digital Library & Electronic database & \url{http://dl.acm.org} \\ \hline
   SCOPUS & Indexing system & \url{http://www.scopus.com} \\ \hline
   Web of Science & Indexing system &  \url{http://webofknowledge.com} \\ \hline
   \end{tabular}
   \end{center}
\end{table}

We selected the above sources due to their recognized effectiveness in supporting systematic studies in software engineering~\cite{Brereton2007571}. 
Starting from our research goal and questions, we constructed a search string to query the selected databases and indexing systems. We deliberately employed a relatively simple search string to capture as many potentially relevant studies as possible. Although we are aware of numerous specific AI techniques, we chose not to include them explicitly for two reasons. First, incorporating only a subset of AI techniques would have introduced arbitrariness and risked biasing the retrieved literature. Second, while including a comprehensive list of AI techniques might have been feasible, such an approach would likely have returned a disproportionately large number of primary studies, rendering the review process unmanageable. Potential threats to validity related to this choice are discussed in Section~\ref{sec:threats}. The final search string was:
\begin{center}
("Artificial Intelligence" OR AI) AND "software architecture*" 
\end{center}
The automatic search for peer-reviewed literature yielded 874 candidate studies. We applied a publication year filter, retaining only studies published between January 2019 and August 2025, corresponding to the period during which this review was conducted. From this initial set, we removed non-research results (e.g., editorials, prefaces) and merged duplicates, resulting in 597 potential primary studies.
Following the selection process proposed by Ali and Petersen~\cite{ali2014evaluating}, we applied inclusion (IC) and exclusion criteria (EC) to titles, abstracts, and keywords. The selection criteria used to filter the search results are as follows:
\begin{itemize}
\item Inclusion criteria
    \begin{enumerate}
        \item Studies describing interplay between AI and SA.
        \item Studies describing application of SA techniques to AI.
    \end{enumerate}
\item Exclusion criteria
    \begin{enumerate}
        \item Studies published as tutorial papers, short papers (less than 4 pages), poster papers, editorials and manuals.
        \item Studies not available as full-text.
        \item Secondary or tertiary studies.
    \end{enumerate}
\end{itemize}
To proceed to the next stage, each study had to satisfy at least one IC and EC. This selection phase involved all authors in key decisions, yielding a set of 62 peer-reviewed studies.
To mitigate potential construct validity biases~\cite{Greenhalgh:2005} and ensure the inclusion of relevant studies from the SA community, we manually screened the proceedings (including companion proceedings) of the International Conference on Software Architecture (ICSA) and the European Conference on Software Architecture (ECSA) from 2019 to 2025, using the same selection criteria. This step added 14 more studies. Subsequently, we conducted closed recursive backward and forward snowballing~\cite{snowball}, but identified no additional sources resulting in a total of 76 studies from the automatic search and selection process. 

\subsection{Data extraction, analysis and synthesis}
To extract and analyse data from the selected primary studies, we applied the grounded theory methodology~\cite{grounded_theory}, which offers a systematic process for decomposing qualitative data, categorizing it, and establishing relationships between categories to uncover emerging themes.

Our analysis involved multiple stages. First, we conducted open coding by reading each study in detail and identifying discrete concepts and observations relevant to the application of AI to SA. These codes were generated inductively, without imposing pre-defined categories, to capture the full diversity of reported practices, techniques, and outcomes.

Next, we grouped and refined the open codes into axial codes, higher-level categories linking related concepts. This stage was carried out iteratively, revisiting both the codes and the primary studies to ensure that categories were well-founded, mutually coherent, and grounded in the source material. Throughout the process, we maintained explicit traceability between codes and their originating studies to support transparency and reproducibility.

This multi-stage application of grounded theory enabled us to rigorously structure and synthesize qualitative evidence, revealing recurring patterns, conceptual relationships, and research gaps in the use of AI within SA.

During the data extraction, we further removed studies from which we could not extract any relevant information obtain a final set of 51 primary studies that are listed in the Primary Studies appendix.


\subsection{Integration of open software architecture challenges}
Concurrently with the SLR, all authors engaged in a qualitative process to revisit and refine the comprehensive list of open SA challenges identified in our peer-reviewed study, which was based on interviews with 32 practitioners~\cite{wan2023software}. To the best of our knowledge, this work is the only peer-reviewed source that explicitly enumerates architectural challenges, or challenges directly related to architecting software systems, which is why we adopted it as the reference framework for our analysis.

We first extracted and consolidated the challenges reported in our previous study, clarifying their scope, removing overlaps, and aligning terminology with our study. This refinement was conducted iteratively by all five authors, with repeated reviews and consensus, building to ensure conceptual clarity and consistency.

To integrate the literature with these empirical challenges, we applied a structured two-step procedure.

First, each topic derived from our grounded theory analysis was systematically mapped to one or more challenges. The mapping was guided by the following criteria: (i) direct evidence, where the primary study explicitly stated that the AI technique addressed the challenge; (ii) inferential evidence, where the reported practice clearly related to the challenge even if not explicitly stated; and (iii) an exclusion principle, where no mapping was made if the relation was speculative or too generic. Mappings were initially performed independently and then consolidated through consensus meetings to reduce subjectivity.

Second, we examined the extent of coverage: AI contributions could fully address, partially address, or fail to address a given challenge. In cases of partial or absent coverage, we formulated new AI-specific challenges (AICHs) that capture the remaining gap. For example, the empirical challenge SACH2 – Architecture Documentation Becomes Obsolete as Software Evolves~\cite{wan2023software} is linked to the AI techniques T5 – AI Knowledge Representation and T8 – AI-Assisted Architecture Recovery and Reverse Engineering. While these techniques partially mitigate the problem, they do not ensure end-to-end traceability and ongoing alignment between architecture documentation and evolving systems. This gap leads to the AI-specific challenge AICH2 – Ensure Traceability and Alignment.

Finally, to strengthen credibility, mappings and gap identifications were iteratively validated in joint author meetings.

Notably, multiple AI practices may contribute to the same empirical challenge, either complementing each other or addressing different aspects of the problem. This rigorous, evidence-driven mapping process, grounded in both the extracted data and our collective expertise in AI and SA research, makes transparent where and why current approaches fall short and directly informs our proposed roadmap for advancing AI4SA.


\section{Review Synthesis: The Current State of Artificial Intelligence in Software Architecture} \label{sec:synt}

In this work, we present a synthesis of the key insights from our SLR, identifying three major thematic clusters that characterize the interplay between AI and SA (Table~\ref{tab:synth}). 
The first cluster, AI for Architecture Design and Decision-Making, and the second, AI for Architecture Evolution and Adaptation, illustrate how AI techniques are applied to enhance architectural design, support informed decision-making, and facilitate continuous system evolution. The third cluster, Architecture for AI Systems, highlights how traditional SA principles and practices are evolving to accommodate the unique requirements of AI-based technologies. Given the scope of our work, in this section we focus our analysis on the first two clusters. However, recognizing the growing importance of this third cluster, we briefly discuss some findings in Section~\ref{sec:dis}.

For each primary study reported in Table~\ref{tab:synth}, we extracted not only the described AI technique and its targeted architectural task, but also any information the authors provided regarding evaluation, validation, or real-world deployment. Where such details were absent, we deliberately refrained from speculative interpretations, restricting our synthesis to what was explicitly reported. To further strengthen the link between research and practice, we later introduce a dedicated subsection highlighting concrete real-world examples drawn directly from the primary studies.

\begin{table*}[htbp!]
\begin{small}
\begin{center}
\caption{Synthesis of studies on the interplay between AI and SA.}
\label{tab:synth}
\begin{tabular}{| p{1cm} | p{2cm} | p{7.3cm} | p{3.3cm} |}
   \hline
\textbf{Contrib.} & \textbf{Cluster} & \textbf{Topic} & \textbf{Study} \\
   \hline\hline 
   
   \multirow{14}{3cm}{\parbox{2cm}{\centering\rotatebox{90}{AI for SA}}} 
   & \multirow{6}{2cm}{C1-AI for Architecture Design and Decision-Making} &
   T1-AI-Assisted Architecture Design & \cite{P2, P16, P20, P21, P36, P38, P42, P47, P48}\\ \cline{3-4}
   & & T2-AI-Assisted Design Pattern Recognition & \cite{P14} \\ \cline{3-4}
    & & T3-AI-Assisted Design Decision Analysis & \cite{P10}  \\ \cline{3-4}
    & &  T4-AI Knowledge Representation & \cite{P6, P8} \\ \cline{3-4}
   & & T5-AI-Assisted Design Decision and Decision Making & \cite{P27, P29, P31, P35, P37, P40} \\ \cline{2-4}
   
   & \multirow{8}{2cm}{C2-AI for Architecture Evolution and Adaptation} & 
   T6-ML Component Adaptation & \cite{P33} \\ \cline{3-4} 
   & & T7-AI-Assisted Architecture Recovery and Reverse Engineering & \cite{P25, P30} \\ \cline{3-4} 
   & & T8-AI-Assisted Resource Management & \cite{P24} \\ \cline{3-4} 
   & & T9-AI-Assisted Performance Estimation & \cite{P11} \\ \cline{3-4} 
   & & T10-AI-Enabled Industrial Automation & \cite{P7} \\ \cline{3-4} 
   & & T11-AI-Assisted Resilience & \cite{P4} \\ \cline{3-4} 
   & & T12-AI-Assisted QAs & \cite{P26, P34, P39}\\ \hline

\end{tabular}
\end{center}
\end{small}
\end{table*}

\subsection{C1 - AI for Architecture Design and Decision-Making}
This cluster groups studies that explore how AI techniques can augment or automate various aspects of architectural design and decision-making. The central idea is that AI can assist architects in navigating the complexity of design choices, pattern selection, and trade-off analysis, thereby enhancing the efficiency and quality of architectural outcomes. Across the included topics, AI is used both as a support tool (e.g., recommending patterns or analysing alternatives) and as a knowledge medium (e.g., representing architectural knowledge explicitly for reuse). 
Collectively, these studies illustrate how AI is beginning to act as both a reasoning partner and a knowledge management medium in architectural design.

\subsubsection{T1 - AI-Assisted Architecture Design}
Research in this topic explores how AI techniques can support the transition from requirements to architecture, automate early architectural decisions, and assist in generating and maintaining architectural artifacts. The studies in this category highlight diverse approaches, ranging from natural language processing (NLP) and graph-based analysis to the exploratory use of large language models (LLMs).

Eisenreich et al. articulate a semi-automated, end-to-end process that generates architecture candidates from requirements, records Architecture Decision Records (ADRs), and applies ATAM-like trade-off analysis with LLM support~\cite{P2}. Although presented as a vision with planned evaluations, the work is notable for proposing tight process integration and for anticipating risks such as model leakage when using reference architectures.

Gustrowsky et al. fine-tune Llama 2 (7B) with QLoRA on a custom, labelled dataset to suggest architecture patterns from requirements, producing both a pattern recommendation and an explanatory rationale~\cite{P16}. On a small, three-pattern scope (MVC, microservices, client–server), the model achieves 70\% accuracy and often provides useful justifications; however, it also exhibits output drift (continuing text past end-of-sequence) and pattern bias (favoring MVC/client–server). 

Rodriguez et al. propose a semi-automated approach based on NLP and data mining to derive architectural responsibilities and Use Case Maps directly from textual requirements~\cite{P21}. Their method identifies functional responsibilities, extracts causal relations, and clusters them into conceptual components, producing a first-cut architectural view. Evaluation across four projects achieved approximately 75\% recall, demonstrating that AI can reduce manual effort in deriving architectural models from requirements.

Arias et al. extend this line of work by applying NLP, graph analysis, and community detection to automate microservice partitioning from functional requirements~\cite{P20}. Their approach further incorporates LLMs such as ChatGPT to generate candidate microservice names, alleviating the need for fully manual decomposition and demonstrating how AI can contribute to early architectural design choices.

The use of LLMs has also been explored in the context of reasoning about architectural patterns. Guerra and Ernst examine the effectiveness of ChatGPT-4 Turbo in understanding and reasoning about the VIPER architecture for iOS applications~\cite{P36}. Using Bloom’s taxonomy as a benchmark, their study shows that LLMs perform well in higher-order cognitive tasks such as evaluating or proposing alternative solutions, while being less reliable in recalling precise architectural structures. This suggests potential for AI to augment creative architectural reasoning, albeit with limitations in accuracy.

Beyond reasoning support, AI has also been employed to generate or extract architectural elements. Fuchß et al. show how LLMs can recover traceability links by extracting component names from documentation and source code, achieving near state-of-the-art performance without intermediate models~\cite{P47}. Similarly, Arun et al. investigate the ability of LLMs to generate serverless architectural components, specifically functions-as-a-service (FaaS)~\cite{P48}. Their findings indicate that LLMs can produce viable components that integrate into existing systems when guided with context-rich prompts, although human review remains necessary to ensure quality.

Finally, Ullah et al. apply transformer-based models, including GPT-3, to identify self-admitted technical debt and software defects from code and comments~\cite{P38}. While not generating architectures per se, this line of work contributes to architectural decision-making by providing insights into design trade-offs and quality degradation.

Collectively, these works suggest that AI techniques, particularly LLMs and NLP, can assist architects in moving from requirements to candidate architectures, though reliability, bias, and evaluation in industrial contexts remain open challenges.

\subsubsection{T2 - AI-Assisted Design Pattern Recognition}
Design patterns provide reusable solutions to recurring design problems, but recognizing their implementation in source code remains difficult, particularly in large or evolving systems. Traditional approaches typically rely on static analysis, Abstract Syntax Trees (ASTs), or graph-based methods, which are costly to apply at scale and require full source code access.
Recent work has investigated the use of programming language models (PLMs) for automating pattern recognition. Pandey et al. conducted a comparative study of four PLMs, OpenAI CodeX, Facebook TransCoder, ACoRA, and CCFlex, on detecting the Singleton pattern in C++ programs~\cite{P14}. Their results show that PLMs can identify Singleton implementations without explicit reliance on naming conventions, suggesting potential for generalization. However, performance varied across models, with some models favouring syntactic cues while others responded more to semantic variations. In addition, rankings of examples were inconsistent across deployable architectural support mechanisms, raising concerns about reliability in industrial contexts. This highlights the need for more robust benchmarks and broader evaluations beyond single patterns and programming languages. 

\subsubsection{T3 - AI-Assisted Design Decision Analysis}
Dhar et al. investigate whether LLMs can support design decision analysis by transforming Architectural Decision Record (ADR) contexts into candidate design decisions, thereby assisting architects in articulating and documenting Architectural Design Decisions (ADDs)~\cite{P10}. Using a corpus of 95 ADRs, the study compares zero-shot, few-shot, and fine-tuned configurations across GPT and T4 families, evaluating outputs with ROUGE, BLEU, METEOR, BERTScore, and manual inspection. Results show that GPT-4 produces coherent, context-appropriate decisions in zero-shot settings but remains sub-human in completeness; GPT-3.5 achieves comparable quality with few-shot prompting; and fine-tuned Flan-T4 attains competitive performance while being feasible for on-premise deployment. The work indicates that LLMs can reduce effort in decision analysis and ADR maintenance, but human oversight is still required for accuracy, traceability, and rationale quality. Open issues include industrial validation, integration into AKM/ADR workflows, explainability of decision rationales, and privacy concerns for cloud-hosted models. Overall, LLMs show promise for augmenting decision analysis but require integration into workflows and stronger empirical validation.

\subsubsection{T4 – AI Knowledge Representation}
Knowledge representation plays a central role in supporting architecture-related decisions for AI adoption. Jahić et al. propose the AI-PAG template, which decomposes drivers for AI adoption into architecturally significant properties such as model training, deployment, or compliance~\cite{P6}. Architects are asked to record existing knowledge, identify adoption stages, and plan next steps, thereby making explicit where knowledge gaps exist. Applied in industry, AI-PAG improved adequacy assessments and increased confidence in AI-related architectural decisions by grounding them in explicit knowledge traces.

Lu et al. address knowledge representation in the context of foundation model–based systems~\cite{P8}. They propose a taxonomy that organizes architectural decisions along three axes: pretraining and adaptation strategies, design options at the model and system level, and responsible-AI-by-design measures. This taxonomy provides a framework for comparing alternatives and highlights trade-offs in cost, accuracy, and trustworthiness. Importantly, it integrates responsible AI concerns into architectural reasoning, ensuring that ethical and governance issues are explicitly considered in design.

Both contributions demonstrate how structured representations can formalize architectural knowledge, but they also reveal the lack of large-scale evaluations in diverse industrial settings

\subsubsection{T5 - AI-Assisted Design Decision and Decision-Making} 
Recent advances demonstrate how AI can assist software architects in making complex design decisions by capturing preferences, reducing decision fatigue, and guiding reasoning in evolving contexts.

Kuviatkovski et al. extend the MOA4PLA tool for product line architecture optimization with interactive modules that incorporate Machine Learning (ML)~\cite{P31}. Their approach learns the decision maker’s preferences during early interactions and later substitutes human input to reduce fatigue. By evaluating different ML algorithms, they show that alternative models outperform multilayer perceptrons in both accuracy and processing time, thereby generating more representative architectural solutions. This highlights the potential of AI to complement human judgment in multi-objective optimization tasks.

Generative AI has also emerged as a decision-support tool. Maranhão and Guerra introduce a prompt pattern sequence approach, where structured prompts guide GPT-based models in producing recommendations aligned with functional and non-functional requirements~\cite{P35}. Evaluated in real-world and fictional projects, this method supports context-aware decision flows while emphasizing the need for human oversight to validate assumptions. A follow-up study applies this technique specifically to microservices, showing that generative AI can reproduce architectural decisions consistent with those taken by industry architects, though with limitations in depth and detail~\cite{P37}.

At a broader knowledge management level, Dhar et al. propose leveraging LLMs to generate and retrieve architectural knowledge, including decision records~\cite{P40}. Their exploratory study demonstrates that GPT-4 and fine-tuned smaller models can generate architectural design decisions from distributed artifacts such as code, diagrams, and ADRs, thus reducing knowledge vaporization and supporting decision traceability.

Other studies broaden the decision-making landscape. Muccini et al. present a framework that combines AI-driven analysis with quality attribute trade-off evaluation to guide architectural choices in complex systems~\cite{P27}. By encoding decision alternatives and their projected impacts, the framework helps architects compare design options systematically. 
Karetnikov et al. explore reinforcement learning for adaptive decision support in software architecture, demonstrating how decision policies can be optimized over time as systems evolve and new operational data becomes available~\cite{P29}. Both approaches highlight the potential of AI to move beyond static decision aids toward adaptive and context-sensitive guidance.

Taken together, these works illustrate how AI can provide both interactive and autonomous support for architectural decision-making, though integration into organizational workflows and concerns around trust and explainability remain open challenges.

\subsection{C2-AI for Architecture Evolution and Adaptation}
This cluster groups studies that leverage AI to support the evolution and adaptation of software architectures over time. Unlike design-time assistance, which focuses on deriving or evaluating architectural structures, this cluster addresses the challenges of keeping architectures aligned with continuously changing requirements, workloads, and operational environments. The reviewed works span a wide range of application domains and techniques, from automated recovery and reverse engineering to resource management, performance prediction, resilience, and refactoring. These studies illustrate how AI can enable continuous alignment between intended architectures and their evolving implementations, a key challenge for long-lived systems.

\subsubsection{T6 - ML Component Adaptation}
Abdullah et al. extend the DEECo model with ML-DEECo, introducing estimators as first-class abstractions for embedding ML into self-adaptive architectures~\cite{P33}. Estimators are declaratively bound to components or ensembles and automatically trained at runtime, enabling predictions such as battery depletion or charging time without manual ML integration.
In a smart farming case study, ensembles of drones used estimators to anticipate resource needs and coordinate charging schedules, which reduced crop damage and improved system resilience. The work demonstrates how ML can be elevated to the architectural level to simplify adaptation, though validation is limited to simulations and a single domain. Future work is needed to validate this abstraction in larger-scale and heterogeneous environments.

\subsubsection{T7 - AI-Assisted Architecture Recovery and Reverse Engineering}
Research on AI-assisted architecture recovery has focused on automating the extraction and evaluation of architectures from existing codebases, with particular attention to improving accuracy, scalability, and developer support.

Dobrean’s work proposes a tool-supported approach for mobile applications that combines SDK-based heuristics with clustering algorithms to detect architectural layers (e.g., MVC, MVVM, MVP) and examine inconsistencies~\cite{P25}. The tool builds topological graphs of code dependencies, checks them against prescriptive architectural rules, and reports issues within CI/CD pipelines. This hybrid detection strategy (SDK inheritance plus clustering) yields higher accuracy than previous methods and provides actionable insights, such as percentage of wrong dependencies, thereby supporting both validation and evolution of architectures.

In parallel, Vazquez et al. present a recommender system that addresses the problem of recovering relevant JavaScript packages from web repositories~\cite{P30}. Their approach integrates meta-search techniques with a ML–based ranking model that learns from features extracted from community projects (e.g., downloads, stars, contributors). By combining retrieval across NPM and general-purpose search engines with pairwise learning-to-rank, the system alleviates information overload and improves package selection accuracy. This contributes to architecture recovery and reverse engineering by supporting technology selection, a crucial step in reconstructing realistic architectures of web systems.

Together, these studies demonstrate how AI can support both structural recovery and component recommendation, but current solutions remain domain-specific and require integration into broader recovery pipelines.

\subsubsection{T8 - AI-Assisted Resource Management}
Hettiarachchi et al. propose an AI-based framework for managing microservice deployments in Kubernetes~\cite{P24}. The system integrates low-latency container provisioning, a metrics server, BiLSTM-based load prediction for proactive autoscaling, and resilience evaluation via chaos experiments.
Evaluated on Azure Kubernetes Service, the framework reduced container deployment times by up to 34\% and achieved more accurate forecasts than ARIMA and Holt-Winters, enabling efficient scaling decisions. While tested only on small-scale deployments, the approach illustrates how deep learning can shift resource management from reactive, rule-based scaling toward proactive and adaptive strategies.
This points toward more autonomous, AI-driven resource management, though generalization to large-scale, multi-tenant clusters remains to be demonstrated.

\subsubsection{T9 - AI-Assisted Performance Estimation}
Berquand et al. present the Design Engineering Assistant (DEA), an AI-based expert system to support early spacecraft feasibility studies~\cite{P11} . The DEA integrates Knowledge Graphs, NLP, and reasoning engines to capture distributed design knowledge and provide performance recommendations during concurrent engineering sessions.
A key feature is explicit uncertainty quantification, allowing users to assess not only performance estimates but also associated confidence levels. Evaluation shows that the DEA improves efficiency by reusing historical data and enhances trust in AI-assisted decision support. Although domain-specific, the approach illustrates how AI can aid performance estimation under uncertainty, a challenge equally relevant to software-intensive systems.
The approach illustrates how AI-based expert systems can support performance estimation under uncertainty, but its transferability to general software architecture contexts has yet to be validated.

\subsubsection{T10 - AI-Enabled Industrial Automation}
Sudharson et al. propose a multimodal AI framework for 
hyperautomation in the context of industrial environments. Hyperautomation extends Robotic Process Automation (RPA) by combining it with AI and ML techniques such as NLP, computer vision, and process mining to scale automation beyond repetitive tasks~\cite{P7}. The framework integrates event-driven architectures, digital process automation, and intelligent business process management to enable organizations to rapidly identify automation opportunities, generate automation artifacts (e.g., bots, workflows), and continuously optimize processes.
The authors evaluate hyperautomation against traditional automation using cost, productivity, and maintenance parameters, showing higher productivity but also increased maintenance costs. They identify several challenges, including interoperability among automation tools, incomplete metrics for assessing ROI, and the need for human oversight to refine auto-generated artifacts.
The study underlines both the transformative potential of hyperautomation and the significant barriers to its sustained adoption.

\subsubsection{T11 - AI-Assisted Resilience}
Canonico et al. propose using reinforcement learning (RL) to strengthen chaos engineering practices~\cite{P4}. Traditional chaos monkeys inject random faults, but this limits systematic coverage. The study investigates whether RL agents can more effectively reveal resilience weaknesses by targeting architectural vulnerabilities.
RL agents (DQN, PPO, VPG) were trained on client–server and peer-to-peer systems, receiving feedback based on system capacity loss. Results showed that RL consistently outperformed random injection in identifying critical faults, with DQN achieving the most stable performance. PPO offered higher potential but less predictability.
This suggests RL can provide more systematic resilience testing, though current evidence is limited to simulated settings.

\subsubsection{T12 - AI-Assisted QAs}
Several works explore how AI can enhance Quality Attributes (QAs) in software systems, particularly in microservices and ML-enabled environments.

Abdullah et al. propose a ML-based framework to improve performance, reliability, and maintainability in microservices applications~\cite{P33}. The approach integrates micro frontends with microservices to reduce communication overhead and response times, and applies ML models for proactive reliability improvement under bursty workloads. Empirical evaluation with case-study applications shows gains in throughput and reduced request failures, suggesting that ML can shift QAs assurance from reactive monitoring to proactive adaptation.

Indykov et al. conduct a multiple-case study of four AI solution providers to investigate quality trade-offs in ML-enabled systems~\cite{P39}. Companies consistently prioritized reliability, resource efficiency, and functional suitability, but faced recurring tensions such as efficiency vs. reliability and accuracy vs. explainability. Tactics included cloud-based deployment, containerization, bias mitigation modules, and human-in-the-loop strategies. The study underscores that AI for QAs must account not only for technical improvements but also for balancing competing quality attributes in practice.

Finally, Sun et al.  focus on QAs-aware service composition, proposing a deep reinforcement learning (DRL) approach for dynamic service selection in cloud environments~\cite{P26}. By modeling QAs such as latency, reliability, and availability, the DRL agent learns optimal composition strategies that outperform heuristic and traditional optimization methods. The work highlights the role of AI in enabling adaptive QAs management under dynamic workloads.

Together, these works reveal how AI can optimize individual QAs  and manage trade-offs, but highlight ongoing challenges in explainability and operational integration.

\subsection{Real-World Applications of Artificial Intelligence in Software Architecture}
While many AI-for-SA approaches remain at the proof-of-concept stage, several of the primary studies included in our review provide evidence of real-world applicability, ranging from test-bed experiments to direct industrial deployments. 

A first group of studies demonstrates clear industrial validation. For example, programming language models for design pattern recognition have been applied within the automotive domain in collaboration with Volvo Cars, showing the potential of AI to support architectural reasoning in safety-critical embedded software~\cite{P14}. Similarly, ML-based performance estimation has been evaluated on industrial aerospace systems, predicting latency and throughput during early design stages~\cite{P11}. Real-world case studies in cyber-physical infrastructures, such as smart grids, demonstrate AI’s role in resilience analysis and failure prediction~\cite{P4}. More recently, multiple case studies with four AI solution providers (including healthcare and autonomous driving) reveal recurring trade-offs in ML-enabled systems, highlighting the tension between efficiency, reliability, and explainability in practice~\cite{P39}. These examples confirm that AI has begun to penetrate domains where architectural quality is paramount, even if solutions remain domain-specific.

A second group of studies provides evidence from realistic, but constrained scenarios. Knowledge-based adequacy assessment frameworks have been validated with industrial partners considering AI adoption, emphasizing decision support for architectural feasibility~\cite{P6}. In mobile applications, AI-based architecture checkers have been prototyped for integration into CI/CD pipelines to detect architectural drift caused by SDK evolution~\cite{P25}. Recommender systems for recovering JavaScript packages have been tested at scale on NPM repositories and GitHub projects, reporting measurable improvements over existing search capabilities~\cite{P30}. Resource management approaches have been validated on Kubernetes test-beds, illustrating concrete benefits for cloud-native microservices~\cite{P24}. Similarly, ML-based methods for improving QAs in microservices~\cite{P34} and for server-less architectural component generation~\cite{P48} have been validated on representative open-source systems, offering applied insights while still short of full industrial adoption.

Finally, a third group consists of studies that remain largely academic or illustrative in nature. Techniques for generating architectural candidates from requirements using LLMs~\cite{P2,P16}, automated microservice name generation~\cite{P20}, and responsibility extraction from textual requirements~\cite{P21} demonstrate promising automation but are evaluated primarily on curated requirement sets or academic benchmarks. Similar constraints apply to experiments on representing decision-makers in product line architectures~\cite{P31}, semantic modelling of architecture decision records~\cite{P29}, and prompt-pattern strategies for guiding architectural decision-making~\cite{P35,P37}. Research on leveraging generative AI for architectural knowledge management~\cite{P40} and for comparing alternative architectural designs~\cite{P36} also remains exploratory, relying on constructed examples rather than longitudinal deployment. Other works illustrate AI’s utility in domain exploration—for instance, taxonomy building for foundation-model systems~\cite{P8} or adequacy frameworks for AI adoption~\cite{P6}—yet their validation remains bounded to academic case studies or concept-driven evaluation.

Across this spectrum, one consistent observation emerges: the degree of real-world grounding varies significantly. At one end, we find industrial validations such as design pattern recognition in the automotive domain (T2~\cite{P14}), knowledge-based adequacy assessments (T5~\cite{P6}), performance estimation in aerospace (T10~\cite{P11}), resilience prediction in cyber-physical systems (T13~\cite{P4}), and quality trade-offs in ML-enabled systems (T14~\cite{P39}). At the other end, many contributions remain primarily academic, including early design automation (T1~\cite{P2,P16,P21}), decision analysis (T3~\cite{P10}), and decision-support methods (T6~\cite{P29,P35,P37,P40}). In between, several approaches demonstrate feasibility on realistic testbeds, such as architecture recovery (T8~\cite{P25,P30}), resource management in Kubernetes (T9~\cite{P24}), and selected QAs studies (T14~\cite{P26,P34}).

\section{AI-Specific Challenges in Software Architecture} \label{sec:ch}
Despite AI’s growing role in SA, its current capabilities still fall short of addressing several enduring and practice-critical challenges. As Ozkaya argues~\cite{ozkaya2023next}, the shift toward AI-enhanced architectural practices must be driven by a clear understanding of existing obstacles and limitations, rather than by a fear of missing out on technological trends.

To make these limitations explicit, we systematically aligned the practices and techniques identified in our SLR with a set of 17 open SA challenges (SACHs) previously derived from 32 practitioner interviews~\cite{wan2023software}.
This mapping not only reveals where AI contributions provide concrete support, but also highlights where solutions remain partial, superficial, or entirely absent. In such cases, we distilled the gaps into a set of AI-specific challenges (AICHs) that mark the frontier for advancing AI-driven SA.

Table~\ref{tab:challenges} summarizes the 17 SACHs, showing their relationships to current AI practices (T) and the resulting AICHs. This provides a rigorous bridge between empirical practitioner needs and the current research landscape, establishing the foundation for our vision of AI4SA (see Section~\ref{sec:met} for details on the mapping procedure).

It is important to note that many underlying software architecture challenges have been studied extensively in software engineering research; our contribution is to examine how these persistent issues manifest in the context of AI support and to identify the AI-specific challenges (AICH1--AICH6) that currently prevent contemporary techniques from resolving them. Importantly, these challenges extend beyond traditional runtime adaptation concerns: whereas research on self-adaptive and self-* architectures has extensively investigated how running systems can adapt in response to uncertainty and changing conditions across technical, human, economic, ethical, and situational drivers, AICH1--AICH6 expose a complementary set of lifecycle-wide intelligence gaps in software architecture practice. These include continuous architectural traceability (AICH2), context-aware reasoning over domain semantics (AICH3), and longitudinal erosion and debt modelling (AICH6). Together, these gaps shift the focus from embedding adaptation capabilities within the system itself toward enabling data-driven, generative, and cross-artifact architectural intelligence that supports design-time reasoning, documentation, analysis, maintenance, and long-term evolution.

Collectively, these six AICHs reveal a clear pattern: while AI techniques have demonstrated value in automating architectural tasks, their contributions remain fragmented, reactive, and often detached from the practical realities of architectural work. Across requirements volatility, documentation, design boundaries, etc., AI methods tend to operate on static inputs, localized perspectives, or abstracted records, leaving fundamental practitioner concerns unaddressed. This synthesis underscores that the frontier of AI for SA is not defined by producing more tools in isolation, but by advancing methods that are adaptive, context-aware, evidence-driven, and tightly aligned with the longitudinal and socio-technical nature of architectural practice. 
These insights establish the foundation for our vision in the next section, where we outline directions for closing the identified gaps and moving toward AI4SA.

\begin{table*}[t]
\begin{footnotesize}
\begin{center}
\caption{Reported SACHs, relationships to current AI practices (T), and resulting AICHs.}
\label{tab:challenges}
\begin{tabular}{|  p{9cm} |  p{1.8cm} |  p{3cm} |}
   \hline
\textbf{SA challenge} & \textbf{AI for SA Topic} & \textbf{AI-specific challenge} \\
   \hline\hline 
   SACH1-Unpredictable Evolution and Changes of Software Requirements & T1, T3, T5 & AICH1-Develop adaptive and continuously updated recommendations \\ 
   \hline
   SACH2-Architecture Documentation Becomes Obsolete as Software Evolves & T4, T7 & AICH2-Ensure continuous traceability and alignment between evolving systems and documentation\\ 
   \hline
   SACH3-Unclear Boundaries Between Architectural Elements & T1, T2 & \multirow{2}{*}{\shortstack{AICH3-Develop context-aware \\ reasoning}} \\ 
   \cline{1-2}
   SACH4-Interdisciplinary Knowledge Required to Lower Coupling and Improve Cohesion & T1 & \\ 
   \hline
   SACH5-Architecture Review Requires a Standard Process, External Expertise, and Tool Support & T3, T5 & AICH4-Incorporate domain-specific expertise and contextual knowledge into AI-supported reviews\\ 
   \hline
   SACH6-Lack of Effective and Apply-to-All Quantitative Measures & T9, T12 & \multirow{4}{*}{\shortstack{
AICH5-Develop adaptive, \\ evidence-based quantitative \\ measures of architectural \\ quality
   }} \\ 
   \cline{1-2}
   SACH7-Automated Architecture Conformance Checking Is Rare & T7 & \\ 
   \cline{1-2}
   SACH8-Obsolete Documentation and Lack of Traceability Hinder Conformance Checking & T7 & \\ 
   \cline{1-2}
   SACH9-Limited Tool Support to Continuously Monitor the Health of Software Architectures & T8, T9, T11, T12 & \\ 
   \hline
   SACH10-Pinpointing Architecture Problems Requires a System-wide Perspective & T7, T9 & \multirow{6}{*}{\shortstack{ AICH6-Integrate long-term \\ evolution and debt \\ management into AI-supported \\ recommendations}} \\ 
   \cline{1-2}
   SACH11-Technical Debts Are Introduced to Software Projects & T7, T1 & \\ 
   \cline{1-2}
   SACH12-Lack of Tool Support for Detecting Architectural Smells & T7, T5 & \\ 
   \cline{1-2}
   SACH13-Unawareness of Correlation Between Code Smells and Architecture Problems & T7, T1 & \\ 
   \cline{1-2}
   SACH14-Lack of Tool Support to Capture and Aggregate Symptoms of Architecture Erosion & T7, T8 & \\ 
   \cline{1-2}
   SACH15-Inadequate Tool Support for Impact Analysis of Architectural Changes & T3, T5 & \\ 
   \hline
\end{tabular}
\end{center}
\end{footnotesize}
\end{table*}

\subsection{AICH1 - Develop adaptive and continuously updated recommendations}
Practitioners consistently reported that the unpredictable evolution and changes of requirements (SACH1) make it infeasible to design an architecture that remains valid over the long term~\cite{wan2023software}. Interviews highlight how shifts in business needs, user groups, and technology stacks can invalidate assumptions behind key quality attributes. Current AI approaches help, but they are largely static.

In T1 (AI-Assisted Architecture Design), for example, AI is used to derive candidate architectures or architectural elements from requirements, e.g., semi-automated requirement-to-architecture flows~\cite{P2}, LLM-based pattern suggestions from requirements~\cite{P16}, automated naming/partitioning for microservices~\cite{P20}, and responsibility extraction from textual requirements~\cite{P21}. In T3 (AI-Assisted Design Decision Analysis), AI transforms architectural knowledge into candidate decisions or supports ADR analysis, e.g., LLMs generating architectural design decisions~\cite{P10}. In T5 (AI-Assisted Design Decision and Decision Making), AI captures preferences and guides trade-offs—e.g., ML-driven proactive decision support~\cite{P27}, learning surrogates of decision makers in search-based PLA~\cite{P31}, and prompt-pattern decision guidance~\cite{P35,P37}. 

While these works generate alternatives or rank options based on historical or static inputs, they typically require full re-runs, prompt re-engineering, or model retraining when requirements change, and they rarely incorporate continuous feedback from evolving system contexts. This mismatch between volatile requirements (SACH1) and AI’s current snapshot-style reasoning motivates AICH1: moving beyond one-time recommendations toward adaptive approaches that continuously refine and update architectural guidance as requirements, contexts, and operational evidence evolve.

\subsection{AICH2 - Ensure continuous traceability and alignment between evolving systems and documentation}
Practitioners repeatedly reported that architecture documentation becomes obsolete as systems evolve (SACH2), making it unreliable as a reference for decision-making~\cite{wan2023software}. One architect noted that documentation was already outdated after only a few sprints, while another described how maintaining updated documentation was seen as extra work with little short-term payoff, leading to systematic neglect. As a result, teams often relied on tribal knowledge or ad hoc communication, which increased architectural drift and created barriers for onboarding and collaboration.

Existing AI approaches partially address this gap. AI knowledge representation (T4) (e.g., \cite{P6,P8}) structures and formalizes architectural knowledge, while AI-assisted architecture recovery and reverse engineering (T7) (e.g., \cite{P25,P30}) reconstructs it from implementation artifacts. However, both remain insufficient: T4 provides static representations that do not evolve with changing systems, and T7 produces reactive, snapshot-style reconstructions rather than continuously synchronized views.

This mismatch crystallizes into AICH2: the need for AI methods that ensure continuous, bi-directional traceability between system evolution and architectural documentation.
Traceability has long been recognized as a fundamental concern in SE, but AI-based approaches such as knowledge representation or reverse engineering remain largely reactive and static, failing to provide the continuous, bi-directional alignment practitioners demand.
Achieving this would mean moving from static or reactive approaches to living documentation that evolves in lockstep with implementation and decision-making, directly addressing practitioners’ concern that once outdated, documentation is almost never updated.

\subsection{AICH3 - Develop context-aware reasoning}
Practitioners emphasized that architectural reasoning often breaks down when boundaries between architectural elements are unclear (SACH3) or when system design requires integrating heterogeneous domains of knowledge to manage coupling and cohesion (SACH4)~\cite{wan2023software}. These challenges highlight the inherently context-dependent nature of architecture: components must be distinguished not only syntactically, but also semantically, and their interactions must reflect domain-specific concerns.

Current AI-assisted design approaches (T1) have made progress in supporting early-stage boundary definition and responsibility assignment. For example, 
AI techniques can generate architecture candidates directly from requirements (e.g., ~\cite{P2, P16, P21}), propose architectural patterns~\cite{P16}, or extract component boundaries and names from code or textual descriptions~\cite{P47, P48}. Similarly, design pattern recognition (T2) has been used to detect recurring structures in source code that may correspond to higher-level abstractions~\cite{P14}.



However, these AI solutions remain detached from the broader system and organizational context. Automatically generated elements (e.g., microservice names~\cite{P20} or component extractions~\cite{P47, P48}) capture structure but miss semantic meaning and domain-driven considerations. Similarly, pattern recognition models~\cite{P14} detect recurring structures without reasoning about appropriateness given quality or organizational constraints. As noted in the state-of-practice survey~\cite{P42}, LLMs still struggle to encode the nuanced domain knowledge required for context-aware reasoning.

This mismatch between what AI currently delivers (boundary suggestions, structural recognition) and what practitioners require (reasoning about semantics, trade-offs, and domain-specific constraints) brings into focus AICH3: the need to develop context-aware reasoning. Future AI systems must not only identify or generate architectural elements, but also situate them within the evolving business, technical, and quality landscape, supporting architects in making decisions that go beyond structural correctness to embrace contextual adequacy.

\subsection{AICH4 - Incorporate domain-specific expertise and contextual knowledge into AI-supported reviews}
Architecture reviews are a critical activity for ensuring system quality, yet practitioners have long emphasized that they require a standard process, external expertise, and substantial tool support (SACH5)~\cite{wan2023software}.

Current AI-assisted approaches address parts of this challenge by helping architects generate, analyse, and document design decisions. For instance, LLM-based models can suggest architectural decisions~\cite{P10}, while other works apply ML to proactive decision-making~\cite{P27, P31}, semantic modelling of ADRs~\cite{P29}, or prompt-pattern strategies to improve traceability~\cite{P35, P37}. 

Despite these advances, AI support for architecture reviews remains narrow and disconnected from the domain expertise human reviewers bring. Tools can accelerate reviews, but rarely embed the contextual knowledge needed to evaluate trade-offs in real business or technical settings. For example, pattern-sequence prompting improves consistency~\cite{P35, P37}, but does not ensure compliance with regulatory or organizational constraints, while LLM-based decision generation~\cite{P10} offers plausible suggestions without grounding them in system realities or stakeholder priorities. This gap reflects a broader limitation: AI techniques rely on abstracted records or training data, but miss the tacit expertise architects apply in practice.

Thus, AICH4 emerges: the need to incorporate domain-specific expertise and contextual knowledge into AI-supported architecture reviews. Meeting this challenge will require hybrid approaches that combine automated analysis with structured elicitation of domain insights, enabling AI tools not only to formalize decisions, but also to evaluate their appropriateness within the specific organizational, technical, and regulatory contexts where architectural reviews take place.

\subsection{AICH5 - Develop adaptive, evidence-based quantitative measures of architectural quality}
Practitioners emphasized the absence of reliable and universally applicable quantitative measures for architectural quality (SACH6), as well as the rarity of automated conformance checking (SACH7), outdated documentation hindering traceability (SACH8), and limited support for continuous monitoring of architectural health (SACH9)\cite{wan2023software}. 

AI-based techniques have started addressing these issues, though only partially. For instance, performance estimation (T9) has been applied to early design phases in domains like aerospace~\cite{P11}, while QAs prediction methods (T12) have leveraged ML to anticipate service-level trade-offs in microservices and ML-enabled systems~\cite{P26,P34,P39}. Architecture recovery techniques (T7) have explored code- and repository-based analysis to support conformance checking~\cite{P25,P30}, and resource management approaches (T8) have introduced centralized AI-driven monitoring~\cite{P24}. Additionally, resilience-oriented solutions (T11) aim to predict failures and improve robustness in socio-technical settings~\cite{P4}.

However, these methods remain fragmented and domain-specific. Performance models rarely generalize across contexts, QAs predictors depend heavily on historical workloads, and recovery tools still struggle with outdated or inconsistent documentation. Likewise, resource management and resilience techniques focus on narrow operational scenarios without integrating into broader architectural evaluation. The resulting AI-specific challenge (AICH5) is to develop adaptive, evidence-based quantitative measures that can evolve with the system, unify performance, resilience, QAs, and conformance perspectives, and provide architects with continuous, actionable insights into architectural health.

\subsection{AICH6 - Integrate long-term evolution and debt management into AI-supported recommendations}
Several practitioner-reported challenges emphasize that architectural problems cannot be diagnosed or resolved in isolation but require a system-wide and long-term perspective. These include the difficulty of pinpointing architectural problems holistically (SACH10), the persistence of technical debt (SACH11), the lack of tools for detecting and correlating architectural smells (SACH12–SACH13), the absence of mechanisms to track erosion symptoms (SACH14), and the inadequate support for analysing the impact of changes (SACH15)~\cite{wan2023software}.

Existing AI techniques only partially respond to these concerns. Architecture recovery and reverse engineering (T7) can expose inconsistencies and violations~\cite{P25,P30}, but such analyses are static snapshots that ignore how problems accumulate as systems evolve. Performance estimation (T9) and resource management (T8) provide localized quality or efficiency metrics~\cite{P11,P24}, yet these remain disconnected from debt-related issues such as erosion or cross-cutting architectural degradation. AI-assisted design (T1) and decision-support techniques (T3, T5) propose alternatives and analyze trade-offs~\cite{P2,P10}, but they treat decisions as discrete events rather than situated within the historical trajectory of the system.

Taken together, these limitations mean that AI currently addresses architectural problems in isolation, focusing on snapshot-based states or narrow quality attributes. What is missing is the ability to integrate longitudinal evidence—linking architectural smells, debt accumulation, and erosion trends—with AI-supported recommendations. This mismatch between the dynamic, debt-laden nature of real systems and the static reasoning of current AI tools defines AICH6: the challenge of embedding long-term architectural evolution and debt management into AI-assisted decision making.
While AICH1 and AICH2 expose the shortfalls of static reasoning in requirement evolution and documentation, AICH6 extends this critique to the broader dynamics of debt accumulation, erosion, and architectural health.

 \section{Mainstream AI-supported tools for SA: intelligence levels and coverage of identified challenges} \label{sec:tools}
While the previous sections synthesized the state of research and distilled six artificial intelligence-specific challenges (AICH1--AICH6), an important complementary perspective is the current industrial landscape. In this section, we examine a set of mainstream tools used in contemporary architectural workflows, characterize the degree to which they embed intelligence support, and relate these capabilities to the practitioner-reported software architecture challenges (SACHs) and the artificial intelligence-specific challenges (AICH1--AICH6) identified in our analysis.
We emphasize that our contribution is a research roadmap derived from evidence synthesis: the tool snapshot in this section serves only to contextualize current industrial maturity and to highlight capability gaps relative to AICH1--AICH6.

\subsection{Illustrative selection and intelligence levels}
The set of tools discussed in this section was identified in March 2026 through a lightweight, illustrative grey-literature analysis informed by our domain knowledge and experience with contemporary architectural practice. The goal was not to produce an exhaustive catalogue or a vendor comparison, but to capture a representative snapshot of widely adopted categories of industrial tooling that architects commonly rely on across the architectural lifecycle, including documentation and decision capture, modelling and visualization, conformance and quality governance, observability and operational intelligence, and cloud architecture guidance. 

For each selected tool, we consulted the information made publicly available by the corresponding vendor, primarily through official product websites and documentation pages, in order to characterise its primary architectural function and advertised intelligence-related capabilities. The resulting snapshot should therefore be interpreted as time-bound, given the rapid evolution of these tools. To derive the Intelligence Support Levels (ISLs), we applied the same grounded, inductive reasoning process described in Section~\ref{sec:met}, using the documented capabilities of each tool as the basis for classification. In addition, although our systematic literature review did not explicitly target large language models as a search term, we included several LLM-based tools to reflect their growing presence in architectural workflows and their practical influence on contemporary software engineering practice.

Table~\ref{tab:tools_sachs} summarizes the tool set, including each tool’s primary architectural function, its Intelligence Support Level (ISL), and its mapping to SACHs. ISL is not a normative ranking, but a conceptual abstraction used to distinguish observable differences in the nature of intelligence support:
\begin{itemize}
\item Level 0: No intelligence support (manual or static tooling).
\item Level 1: Rule-based automation or template-driven support.
\item Level 2: Assisted intelligence (e.g., structured guidance, heuristic suggestions).
\item Level 3: Machine learning-driven analytics (e.g., anomaly detection, prediction).
\item Level 4: Generative or agentic intelligence (e.g., large language model-based reasoning, automated remediation).
\end{itemize}
Classification was based on documented functional capabilities (e.g., anomaly detection, generative remediation, predictive analytics) rather than vendor claims. The purpose of ISL is to reveal systemic maturity gradients across the architectural lifecycle, not to compare individual tools.

The mapping between tools and SACHs captures partial mitigation at the workflow level rather than complete resolution. A tool is associated with a SACH when its documented capabilities directly support one dimension of the challenge (e.g., conformance checks for SACH7 or anomaly detection for SACH6). The goal is not to claim that mainstream tools solve these challenges, but to assess whether current industrial intelligence reduces the systemic gaps highlighted by AICH1--AICH6.

\begin{table*}[t]
\begin{footnotesize}
\begin{center}
\caption{Mainstream Tools, ISL, and mapping to SACHs}
\label{tab:tools_sachs}
\begin{tabular}{| p{4cm} | p{5cm} | p{.3cm} | p{4.3cm} |}
\hline
\textbf{Tool} & \textbf{Primary Architectural Function} & \textbf{ISL} & \textbf{SACHs} \\
\hline\hline

Atlassian Jira + Confluence (AI/Rovo) 
& Documentation, coordination, ADR support 
& 4 
& SACH1, SACH2, SACH5, SACH8 \\
\hline

GitHub Copilot 
& Code-level architectural assistance, refactoring, review 
& 4 
& SACH1, SACH3, SACH5, SACH11, SACH13 \\
\hline

GitLab Duo Agent Platform 
& SDLC orchestration and governance 
& 4 
& SACH1, SACH7, SACH11, SACH14 \\
\hline

JetBrains AI + Junie 
& IDE-based architectural refactoring and PR automation 
& 4 
& SACH1, SACH11, SACH13 \\
\hline

SonarQube (AI CodeFix) 
& Architecture governance and quality gates 
& 2 
& SACH7, SACH11, SACH12, SACH13 \\
\hline

Snyk Agent Fix 
& Security remediation and governance 
& 4 
& SACH11, SACH12, SACH13 \\
\hline

Postman Agent Mode 
& API design validation and lifecycle governance 
& 4 
& SACH5, SACH7, SACH9, SACH15 \\
\hline

Swagger Studio (AI) 
& API specification design and governance 
& 2 
& SACH5, SACH7 \\
\hline

Lucidchart + Lucid AI 
& Diagram generation and architectural visualization 
& 2 
& SACH2, SACH3 \\
\hline

Miro AI 
& Collaborative modelling and workshop synthesis 
& 2 
& SACH2, SACH3 \\
\hline

Enterprise Architect (Sparx) 
& Model-based architecture engineering 
& 1 
& SACH2, SACH7 \\
\hline

Structurizr 
& Architecture-as-code and documentation synchronisation 
& 1 
& SACH2, SACH8 \\
\hline

Mermaid 
& Diagram-as-code rendering 
& 1 
& SACH2 \\
\hline

adr-tools 
& ADR lifecycle management 
& 1 
& SACH2, SACH8 \\
\hline

Datadog (Watchdog + Bits AI) 
& Observability, anomaly detection, incident support 
& 4 
& SACH6, SACH9, SACH10, SACH14 \\
\hline

Dynatrace (Davis AI) 
& Causal root-cause analysis and telemetry reasoning 
& 3 
& SACH6, SACH9, SACH10 \\
\hline

New Relic AI 
& AIOps and telemetry analysis 
& 3 
& SACH6, SACH9 \\
\hline

Splunk AI Assistant 
& Query generation and operational analytics 
& 2 
& SACH6, SACH9 \\
\hline

Azure Copilot + Advisor 
& Cloud architecture governance and operational guidance 
& 4 
& SACH1, SACH6, SACH9, SACH15 \\
\hline

AWS Well-Architected Tool + Amazon Q 
& Architecture governance and SDLC assistance 
& 3 
& SACH5, SACH6, SACH15 \\
\hline

Gemini Cloud Assist + ADC 
& Cloud architecture template generation and optimization 
& 4 
& SACH1, SACH5, SACH6, SACH15 \\
\hline

\end{tabular}
\end{center}
\end{footnotesize}
\end{table*}

\paragraph{Copilot and IDE-embedded assistants}
Tools such as GitHub Copilot, JetBrains AI, and GitLab Duo embed generative and agentic capabilities directly into development workflows. These systems primarily support requirement-driven change implementation (SACH1), local refactoring and boundary clarification (SACH3), and remediation of technical debt symptoms (SACH11--SACH13). While they reduce manual effort and accelerate change propagation, their reasoning remains code-centric and workflow-bounded. They do not aggregate erosion signals system-wide (SACH14), nor do they connect local refactorings to architectural intent or long-term design trajectories (SACH10).

\paragraph{Static analysis and governance platforms}
Tools such as SonarQube and Snyk provide rule-based enforcement, conformance checking (SACH7), and automated remediation of quality and security issues (SACH11--SACH13). These platforms strengthen architectural discipline through quality gates and policy enforcement. However, their intelligence is largely local and rule-driven. They do not provide lifecycle-wide architectural reasoning, adaptive metric evolution (SACH6), or longitudinal debt aggregation (SACH14).

\paragraph{Documentation, modelling, and architecture-as-code tools}
Documentation platforms, ADR managers, and modelling tools (e.g., Lucidchart, Miro, Enterprise Architect, Structurizr, Mermaid) mitigate SACH2 and SACH8 by improving representation, traceability discipline, and structural clarity (SACH3). Despite recent AI-assisted drafting features, their support remains artefact-centric. These tools improve syntactic consistency but do not ensure semantic alignment between evolving code, runtime evidence, and architectural rationale. Continuous, bidirectional traceability (AICH2) is therefore not systematically achieved.

\paragraph{API lifecycle and governance tooling}
API-centric platforms (e.g., Postman, Swagger Studio) strengthen structured review processes (SACH5), contract validation, and conformance (SACH7), and partially assist with change impact reasoning (SACH15) within bounded API scopes. However, their intelligence remains layer-specific and does not generalize to cross-layer architectural reasoning or system-wide erosion modelling (SACH10, SACH14).

\paragraph{Observability and AIOps platforms}
Observability and telemetry-driven tools (e.g., Datadog, Dynatrace, New Relic, Splunk) exhibit the highest density of ML-driven intelligence. They strongly address quantitative measurement and continuous monitoring (SACH6, SACH9) and partially support system-wide problem localization (SACH10). Nonetheless, these capabilities are operationally siloed. Runtime insights are rarely integrated into architectural decision repositories or design models, and longitudinal erosion modelling (SACH14) remains weakly supported.

\paragraph{Cloud architecture governance and advisory tools}
Cloud assistants and well-architected frameworks (e.g., Azure Copilot, AWS Well-Architected Tool, Gemini Cloud Assist) provide structured review guidance (SACH5), optimization recommendations (SACH6), and bounded change-impact support (SACH15). Their intelligence is typically infrastructure-centric and checklist-oriented. While generative capabilities assist with template generation and deployment adaptation (SACH1), they do not integrate cross-layer architectural reasoning or lifecycle-wide traceability (SACH2, SACH14).

\subsection{Mainstream AI-supported tools coverage of SACHs}
A cross-category inspection of tool capabilities reveals uneven intelligence distribution across the 17 SACHs. Rather than uniform architectural support, intelligence embedding concentrates in operational and governance domains, while systemic and longitudinal concerns remain weakly addressed.

\paragraph{Operational and quantitative challenges (SACH6, SACH9)}
The highest intelligence density appears in quantitative measurement and continuous monitoring. Observability and AIOps platforms provide mature ML-driven analytics, anomaly detection, and predictive capabilities at runtime scale. However, these operational insights are typically siloed and rarely integrated into architectural knowledge repositories or decision models.

\paragraph{Governance and conformance (SACH5, SACH7)}
Review processes and conformance enforcement benefit from structured governance tooling and rule-based automation. While these mechanisms strengthen policy compliance, they remain predominantly checklist-oriented and local. Holistic architectural reasoning and cross-artifact synthesis are generally absent.

\paragraph{Documentation and traceability (SACH2, SACH8)}
Documentation platforms, modelling tools, and architecture-as-code approaches provide broad artefact-level coverage. Despite this density of support, traceability remains largely syntactic. Continuous, semantic alignment between evolving implementations, architectural rationale, and runtime behaviour is not systematically achieved.

\paragraph{Requirement evolution and boundary management (SACH1, SACH3)}
Generative assistants and refactoring support reduce the cost of implementing requirement-driven changes and clarifying structural boundaries. However, these capabilities operate in reactive, prompt-driven modes and do not provide adaptive architectural recommendations grounded in longitudinal system evidence.

\paragraph{System-wide reasoning and evolution (SACH10, SACH14, SACH15)}
The weakest coverage appears in challenges requiring cross-layer, lifecycle-wide reasoning. Although tools detect localized anomalies, technical debt, or bounded impact effects, no mainstream solution aggregates erosion signals over time or embeds architectural history into forward-looking recommendations. Longitudinal architectural intelligence remains largely absent.

\paragraph{Technical debt and degradation (SACH11--SACH13)}
Static analysis and automated remediation agents moderately address localized debt and smell detection. Nevertheless, aggregation and interpretation of debt at the architectural level remain limited, reinforcing the gap between local issue resolution and systemic architectural health modelling.

\subsection{Implications of Mainstream Tool Coverage for AICHs and the AI4SA Vision}

The coverage analysis shows that mainstream AI-supported tools primarily deliver task automation within bounded workflows (e.g., documentation drafting, code refactoring, telemetry analysis, and checklist-based governance). Despite these benefits, they largely fail to provide context-aware architectural reasoning and continuous traceability across the architectural lifecycle, i.e., they do not systematically connect architectural intent and decisions to evolving implementations and runtime evidence.

First, the moderate yet reactive support for SACH1 confirms AICH1. Generative copilots and cloud assistants accelerate architectural changes triggered by evolving requirements, but they operate in episodic, prompt-driven modes. They do not continuously adapt recommendations based on runtime evidence or architectural drift. The industrial landscape therefore reflects the same “snapshot-style reasoning” limitation identified in the SLR, strengthening the case for adaptive, continuously updated architectural intelligence.
Second, the widespread but artefact-level coverage of SACH2 and SACH8 validates AICH2. Documentation tools, ADR managers, and architecture-as-code platforms improve syntactic consistency, yet none ensures continuous, bidirectional alignment between implementation, runtime signals, and architectural knowledge. The observed fragmentation across tooling categories confirms that traceability remains tool-assisted but not intelligence-integrated.
Third, the limited and structurally narrow support for SACH3 and SACH4 reinforces AICH3. While boundary clarification is assisted through diagramming tools and code-level AI, none of the surveyed tools performs context-aware reasoning that integrates domain semantics, organizational constraints, and quality trade-offs. This gap mirrors the research-level limitation of structurally correct but context-agnostic AI reasoning.
Fourth, the partial coverage of SACH5 through governance tools and copilots confirms AICH4. AI accelerates review preparation and documentation generation, but it does not embed domain-specific expertise or regulatory knowledge into the reasoning process. Industrial tooling therefore validates the need for hybrid human–AI collaborative models grounded in contextual expertise.
Fifth, the strong density of tools addressing SACH6–SACH9 provides partial validation of progress toward AICH5, yet also exposes its incompleteness. Observability and AIOps platforms deliver mature ML-driven analytics, anomaly detection, and QAs prediction. However, these capabilities remain operationally siloed and are rarely integrated into unified architectural quality models. The absence of cross-layer, adaptive, and evidence-based architectural metrics confirms that quantitative reasoning remains fragmented.
Finally, the weak coverage of SACH10 and SACH14–SACH15 directly validates AICH6. Although tools detect technical debt, conformance violations, or runtime anomalies, none aggregates longitudinal erosion signals or embeds historical architectural trajectories into AI-supported recommendations. Industrial AI remains short-term and local, rather than evolution-aware and systemic.

Overall, the observed distribution of intelligence across tooling categories indicates that mainstream AI support remains predominantly local and reactive. Systemic capabilities such as lifecycle-wide traceability, context-aware architectural reasoning, and longitudinal erosion modelling remain weakly represented. This structural gap empirically reinforces the necessity of AICH1–AICH6 and the foundational AI4SA research pillars.

\section{The Road Ahead: Charting a Vision for Artificial Intelligence-Driven Software Architecture} \label{sec:roa} 

The analysis of AICHs in Section~\ref{sec:ch} highlights that, while AI has begun to transform certain aspects of SA, its contributions remain fragmented, often reactive, and insufficiently grounded in real-world practice. To move from isolated prototypes toward trustworthy and impactful adoption, a forward-looking research agenda is required. In this section, we chart such a vision by articulating six strategic Pillars (Ps) for advancing AI-driven SA~\cite{DBLP:conf/fose-ws/Lo23}. Each pillar builds directly on the gaps identified in our SLR and the practitioner-reported challenges, while also addressing broader concerns of trustworthiness, scalability, and industrial applicability. Together, these pillars form a roadmap that connects current research efforts with the long-term goal of making AI a reliable and integral partner in architectural design, evaluation, and evolution. This roadmap reorganizes the six strategic pillars into a sequence of dependent Phases (PHs) depicted by Table~\ref{tab:phases_pillars}.

\begin{table*}[htbp!]
\begin{small}
\begin{center}
\caption{Phased roadmap and corresponding pillars.}
\label{tab:phases_pillars}
\begin{tabular}{| p{5cm} | p{9.3cm} |}
   \hline
\textbf{Phase} & \textbf{Pillar} \\
   \hline\hline
   \multirow{3}{9cm} {PH1--Laying the foundation}
   & P1--Establish living architectural knowledge \\ \cline{2-2}
   & P2--Create benchmarks and datasets \\ \cline{2-2}
   & P3--Develop evidence-based quantitative metrics \\
   \hline
    PH2--Developing the core intelligence
   & P4--Build self-evolving architectural intelligence \\
   \hline
   \multirow{2}{9cm}{PH3--Integration, validation, and adoption}
   & P5--Implement human--AI collaborative architecting \\ \cline{2-2}
   & P6--Conduct industrial validation and drive adoption \\
   \hline
\end{tabular}
\end{center}
\end{small}
\end{table*}

\paragraph{PH1--Laying the foundation}
Before an intelligent system can be created, the foundational infrastructure for data, measurement, and evaluation must be established. This phase combines the core ideas of the following Ps.
    \begin{itemize}
        \item Establish living architectural knowledge. The first step is to develop the {tooling and infrastructure} that creates a "living architectural knowledge" base for an existing software project. This involves building the technology (like knowledge graphs and data pipelines) to integrate and synchronize the project's vast data sources, such as requirements, code, and runtime telemetry.
        \item Create benchmarks and datasets. To support all future work, the community must prioritize creating standardized ``benchmarks and datasets". This provides the ground truth needed to train AI models and consistently measure their performance.
        \item Develop Evidence-based quantitative metrics. Simultaneously, the community must define and develop the "evidence-based quantitative" metrics to guide the AI's reasoning. This involves moving beyond simple code metrics to create reliable measures for complex qualities like maintainability and resilience.
    \end{itemize}

\paragraph{PH2--Developing the core intelligence} With the foundational data and metrics in place, the next step is to build the core AI engine. Using data from Phase 1, research can focus on creating the "self-evolving architectural intelligence". This involves developing AI models (like continual learners or reinforcement learning agents) that can perform a three-stage loop: continuous monitoring, synthesis of new information, and proactive decision-making.

\paragraph{PH3--Integration, validation, and adoption} The final phase focuses on making it usable, trustworthy, and effective in the real world.
\begin{itemize}
        \item Implement human-AI collaborative architecting. The AI "brain" must be integrated into a structured workflow that empowers, rather than replaces, the human architect. This involves designing interfaces and interaction patterns that establish the AI as the "Chief Analyst" and the human as the "Chief Strategist". A key focus is ensuring the AI's recommendations are explainable and trustworthy.
        \item Conduct industrial validation and drive adoption. Finally, the complete human-AI system must be validated in real-world industrial settings. This involves embedding the tools into architects' daily workflows to gather evidence of their effectiveness and iterating based on feedback to ensure the solutions are scientifically rigorous and practically relevant.
\end{itemize}

\subsection{The Self-Evolving Architectural Intelligence}
Current AI applications for Software Architecture remain largely episodic and static. 
Architectural reasoning (AICH1, AICH6) and documentation (AICH2, AICH5) are often treated as isolated, reactive tasks, resulting in fragmented and unsystematic practices. 
To address this limitation, we envision not a collection of disconnected tools, but a unified architectural brain that evolves in tandem with the software system it governs. 
This brain operates through a continuous three-stage feedback loop: continuous monitoring, living architecture knowledge, and quantitative reasoning \& proactive decision-making.

\paragraph{Continuous monitoring} The foundation of architectural intelligence is holistic awareness. 
The AI system must continuously monitor and assimilate signals across the entire system lifecycle, including but not limited to runtime telemetry (performance metrics, error rates, resource utilization), version control history (code commits, pull requests, branching patterns), requirements artifacts (user stories, specification documents), developer communications (discussion boards, chat logs, meeting transcripts), as well as architectural decision records and accumulated technical debt.
By creating such a comprehensive sensory input, the AI addresses persistent challenges of information silos and outdated documentation, thereby providing a real-time pulse of the project.

\paragraph{Living Architecture Knowledge} The goal of this stage is to construct a unified knowledge base or graph (synthesis) that integrates architectural representations from the code repository with raw monitoring data, thereby enabling subsequent reasoning and decision-making. 
Ideally, when a monitoring event occurs—for example, a runtime telemetry alert—the system should first localize the relevant architectural component and then trace it further down to the corresponding functions or files in the codebase. 
Similarly, when requirements change, the system should map these modifications onto architectural adjustments and, in turn, onto the specific code artifacts they affect. 
Achieving this capability requires not only a robust knowledge infrastructure capable of representing large-scale architectures without becoming intractable, but also a sophisticated mapping mechanism that treats architecture as the intermediate layer bridging monitoring data, requirements, documentation, and the codebase. 
This mechanism is inherently complex, and the mappings highlighted here—such as linking monitoring data with the code repository, aligning software requirements with architectural models, connecting architectural models with documentation, and maintaining bidirectional links between architecture and code—are illustrative rather than exhaustive. 

\paragraph{Quantitative Reasoning \& Proactive Decision-Making}  Building on monitoring and synthesized knowledge, the AI engages in evidence-based reasoning that informs architectural choices. 
It can simulate trade-offs by quantitatively modelling the impact of proposed changes (e.g., requirement or architectural) on quality attributes such as performance, cost, maintainability, and security. 
It can also predict architectural drift by detecting when code-level modifications begin to violate architectural principles or accumulate technical debt, tracing the ripple effects of local changes on the global architecture. 
Furthermore, it can generate recommendations by offering actionable solutions, such as targeted refactorings to address performance bottlenecks or updates to requirement documents to reflect recent code changes~\cite{chang2025slicemate}. 
Together, these capabilities transform AI from a passive evaluator into a proactive collaborator. 

Integrated into the loop of monitoring, synthesis, and reasoning, they enable continuous architectural evolution grounded in real-time evidence. 
However, to realize this vision, four practical challenges need to be addressed.

\paragraph{Reliable Mechanisms} Whether in mapping, reasoning, or decision-making, effective support requires LLMs or other AI techniques capable of making accurate decisions, handling large-scale code repositories, and continuously learning from feedback. 
Advanced approaches such as multi-agent coordination~\cite{he2025llm} or reinforcement learning~\cite{sutton1998reinforcement} may be leveraged to ensure both reliability and adaptability throughout long-term software evolution. 
Moreover, program slicing~\cite{chang2025slicemate} offers a complementary technique for localizing code-related issues and improving the precision of architectural analysis.

\paragraph{Reliable Architecture Modelling}
A persistent challenge is determining how to represent software architecture effectively in a knowledge graph. 
How can we ensure that models generated through static analysis or LLMs faithfully capture the true architecture rather than partial or distorted views? 
Addressing this issue requires novel techniques for automated model extraction, validation, and continuous refinement, ensuring that architectural representations remain accurate and trustworthy. 

\paragraph{Actionable Quantitative Metrics}
For reasoning to be effective, AI systems need a richer set of quantitative metrics that can be continuously monitored and integrated into the synthesis process. 
The community may move beyond simple code metrics to establish new measures that rigorously capture complex quality attributes such as evolvability, resilience, and modularity. 
Developing validated, actionable metrics that reflect real architectural concerns is therefore a crucial direction for future research.

\paragraph{Scalable and Cost-Effective Analysis}
Modern software repositories are massive and highly dynamic, with thousands of files changing daily. 
The key challenge is to accurately and efficiently detect architecturally significant changes amidst this noise, while controlling the computational cost and latency of continuous analysis—particularly when leveraging large-scale models such as GPT-5. 
Developing efficient, architecture-aware analysis techniques for large-scale repositories is therefore essential. 
\\\\
To make these research directions more concrete, we provide a few brief illustrative examples of operationalisable capabilities. First, LLMs can support living architectural knowledge by continuously recovering and maintaining traceability links among requirements, ADRs, code changes, and runtime signals, directly addressing SACH2 and AICH2. Second, adaptive decision support can move beyond snapshot-style recommendations by updating trade-off and refactoring suggestions as requirements and system behaviour evolve, aligning with SACH1 and AICH1. Third, in hyperautomated industrial environments, architectural governance mechanisms are needed to manage interoperability, lifecycle cost, and quality impacts of auto-generated automation artefacts, thereby motivating research connected to long-term evolution and debt (AICH6). These examples illustrate feasible engineering outcomes implied by our challenges, rather than prescribing specific tool solutions.
These directions invite research on how to evaluate such capabilities in realistic ecosystems, for example through longitudinal studies of traceability freshness, decision latency, and erosion/debt trajectories.

\subsection{Human-AI Collaborative Architecting}
Architectural decision-making is also socio-technical: while AI can process vast data and generate complex analyses, it cannot fully account for organizational priorities, cultural contexts, or long-term strategic goals. 
Current ``human-centered'' approaches often stop at vague principles, offering little guidance on how architects and AI systems should concretely collaborate in practice (AICH3, AICH4). 
To move beyond this gap, the next stage of research must establish explicit collaboration patterns that enable mutual trust and role clarity between humans and AI. 

The envisioned model casts AI in the role of a chief analyst. 
In this capacity, AI systems manage the ``what'' and the ``how'': providing data-driven evidence, generating risk assessments, simulating alternative scenarios, and conducting trade-off analyses. 
Their primary strength lies in managing the complexity and scale of modern software repositories, where human analysis alone would be prohibitively time-consuming. 
For such contributions to be trustworthy, however, explainability is indispensable: every AI recommendation must be traceable to its underlying data and reasoning process, ensuring that architects can scrutinize and validate its outputs. 
In parallel, architects retain the role of chief strategist, responsible for the ``why.'' 
Their decisions integrate AI’s analyses with implicit contextual knowledge, such as business objectives, team expertise, organizational culture, and long-term planning horizons. 
Human oversight emphasizes critique and guidance: architects can interrogate AI conclusions, supplement them with domain expertise, or reprioritize objectives (e.g., directing the AI to prioritize scalability over short-term cost reduction). 

This division of roles transforms “human-centered” from a slogan into a structured workflow. 
By clearly delineating responsibilities, it ensures that architectural processes remain both computationally rigorous and strategically grounded. 
In this way, AI becomes not a black-box oracle, but a trusted collaborator whose analytical power is balanced by human judgment, enabling more robust and context-sensitive architectural evolution.

\subsection{Benchmarks, Datasets, and Industrial Adoption} 
The credibility and adoption of AI-driven SA will ultimately depend on rigorous evaluation and industrial validation. A recurring weakness in the current landscape is the lack of standardized benchmarks, longitudinal datasets, and reproducible evaluation protocols. Many of the existing techniques have demonstrated feasibility in controlled settings or open-source testbeds, but very few have been validated at industrial scale (Section~\ref{sec:synt}). This not only limits confidence in their scalability and robustness, but also impedes fair comparison across competing approaches. To move forward, the community must prioritize the creation of benchmarks and datasets that reflect the full lifecycle of architectural work, including requirements, decision records, implementation artifacts, and runtime evidence. Evaluation protocols should explicitly assess generalizability, robustness to drift, and impact on human decision-making. Synthetic data generation can help mitigate the scarcity of curated architectural datasets, while standardized benchmarks can ensure comparability across competing techniques. Equally important is validation in real-world industrial settings. Embedding AI solutions into architects’ daily workflows, for instance, as plug-ins for widely used development tools, would provide direct evidence of their effectiveness in practice. Close collaboration with industry can also ensure that evaluation criteria reflect real architectural concerns such as long-term maintainability, regulatory compliance, and team collaboration. By combining reproducible evaluation with industrial adoption, the field can ensure that advances on AICH1–AICH6 translate into solutions that are both scientifically rigorous and practically relevant.

\section{Discussion} \label{sec:dis} 
This study shows that AI has already begun to reshape several aspects of software architecture, yet in ways that are fragmented and uneven. Existing contributions typically concentrate on automating well-bounded tasks such as boundary detection, documentation, or performance prediction. While these provide efficiency gains, they rarely address the broader contextual and long-term concerns highlighted by practitioners. Our synthesis of AI-specific challenges (AICH1--AICH6) demonstrates that current techniques are largely reactive, producing static reconstructions or one-off recommendations, whereas architects require operationalisable architectural capabilities that evolve alongside their systems, integrate domain knowledge, and maintain trustworthiness over time. This gap helps explain why industrial validation remains confined to a few areas, such as performance estimation, resilience, or quality-of-service optimization, while most design automation and decision-support approaches remain at the academic prototype stage.

\subsection{Positioning AI4SA with respect to previous roadmaps}
The research agenda for AI4SA advanced in this work is situated within the lineage of established roadmap contributions, particularly Kramer and Magee's vision of self-managed systems~\cite{garlan2014software} and Garlan's synthesis of the architectural discipline~\cite{garlan2014software}. These foundational works articulated the vision of intelligent architectures capable of self-configuration, self-healing, and self-tuning, primarily through explicit architectural models and feedback control loops (e.g., Sense--Plan--Act or MAPE-K) to manage uncertainty and QAs at runtime. This tradition represents a major milestone in operationalising architectural intelligence.

Our work does not challenge this lineage; rather, it examines how recent advances in contemporary AI reshape the scope and nature of architectural intelligence. Whereas earlier self-* approaches concentrated on runtime adaptation through programmed control strategies, the studies synthesised in this review increasingly target design-time reasoning and lifecycle-wide decision support. This shift reflects a broader technological evolution: AI is no longer confined to executing predefined adaptation strategies but is now capable of synthesising unstructured artefacts (e.g., requirements, documentation, code repositories, discussion traces) to assist architectural cognition.

Importantly, the progress achieved over the past two decades in self-adaptive research should be interpreted as a maturation of runtime autonomy and QAs management. The differences highlighted by our synthesis therefore reflect a complementary focus: while the self-* tradition has strengthened model-based feedback control at runtime, AI4SA shifts attention toward lifecycle-wide architectural reasoning needs such as traceability, boundary clarification, trade-off synthesis, and long-term evolution support.

Our contribution is an evidence-based synthesis grounded in a systematic literature review of 51 primary studies and explicitly aligned with empirically derived practitioner challenges. By mapping 12 identified AI usage patterns (Ts) against 17 practitioner-reported challenges (SACHs), we provide a structured assessment of how contemporary AI engages with long-standing architectural concerns. Many of these concerns echo themes in earlier roadmaps, for example, technical debt and erosion accumulation, yet our analysis reveals persistent intelligence gaps when modern AI techniques are applied.

\begin{figure}[!h]
\centering
    \includegraphics[width=\linewidth]{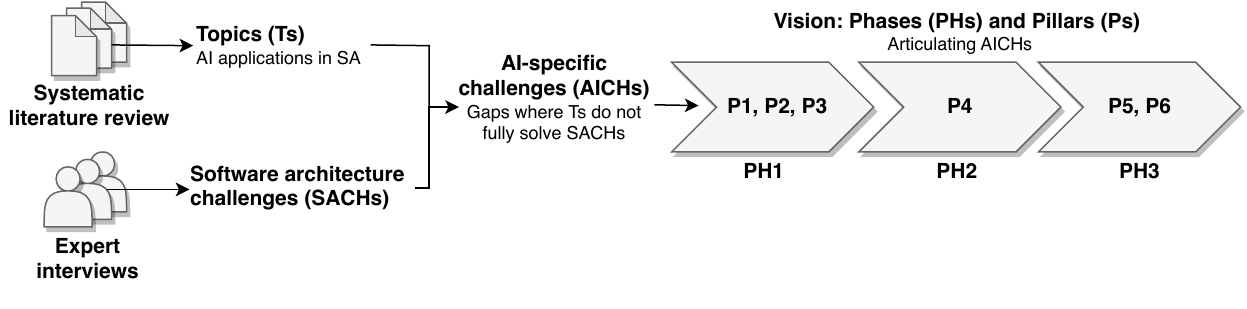}
    \caption{Terminology introduced in this study and relations among the terms.}
    \label{fig:ter}
\end{figure}

The terminology introduced in this study (Ts, SACHs, AICHs) should be understood as synthesis abstractions rather than competing conceptual frameworks. The Ts emerged inductively from grounded coding of contemporary AI-driven architectural studies and classify how AI is currently operationalised. The SACHs anchor the roadmap in empirically observed practitioner needs. The AICHs constitute the principal research frontier: they identify recurring reasoning barriers that prevent current AI techniques from fully addressing established architectural challenges. This is illustrated in Figure~\ref{fig:ter}. For instance, AICH2 (continuous lifecycle-wide traceability), AICH3 (context-aware reasoning over architectural semantics), and AICH6 (integration of longitudinal erosion and debt evidence) highlight limitations specific to contemporary AI paradigms. These challenges are not reformulations of self-adaptive concerns, but intelligence gaps that arise when generative and data-driven AI is applied to architectural tasks beyond runtime control.

In this sense, AI4SA should be interpreted as an evolution of the self-adaptive systems vision in the era of foundation models. Whereas prior roadmaps operationalised architectural intelligence through model-based runtime feedback loops, our findings motivate research on unified, socio-technical, lifecycle-wide architectural intelligence that supports both human decision-making and long-term evolution. This positioning directly addresses our research questions by clarifying (i) how AI is currently leveraged in architectural practice, (ii) where it remains insufficient relative to practitioner needs, and (iii) which intelligence barriers must be overcome to realise the next stage of architectural maturity.

\subsection{Implications for research and practice}
The implications for research are twofold. First, mapping AI contributions against practitioner challenges highlights concrete priorities: adaptive recommendation systems that evolve with requirements (AICH1), living documentation and traceability mechanisms (AICH2), context-aware and explainable reasoning (AICH3--4), evidence-based quality measures (AICH5), and support for long-term evolution and debt management (AICH6). Second, progress in these areas requires more than technical advances in AI models; it depends equally on stronger methodological foundations and evaluation practices. Reinforcement learning or uncertainty-aware inference may enable adaptation, but their success hinges on validation in realistic software ecosystems rather than small-scale testbeds. Similarly, explainability and human--AI collaboration are not optional: unless architects can interrogate, critique, and refine AI-generated suggestions, adoption in safety-critical or regulated domains will remain elusive.

From a practitioner perspective, the findings underscore both the potential and the limitations of current applications. The surveyed techniques suggest that manual burdens can be reduced, documentation kept more accurate, and quantitative trade-offs analysed more systematically. Yet without contextual grounding and continuous traceability, such architectural support mechanisms  struggle to earn trust and remain peripheral in practice. Where stronger evidence does exist, in aerospace, automotive, and cyber-physical systems, it shows that AI can add value when problems are well defined and quality attributes are measurable. More general architectural activities, such as early design or strategic decision analysis, remain poorly supported. This points to a pressing need for AI solutions that embed seamlessly into development workflows, respect organizational and regulatory constraints, and act not as opaque black boxes but as transparent collaborators. For example, embedding LLM-based traceability agents into CI/CD pipelines could provide architects with continuously updated, inspectable links between requirements, code, and architectural decisions, addressing both efficiency and trust concerns.

\subsection{Broader research context}
These observations also resonate with broader debates in software engineering and artificial intelligence. The challenges of explainability, trustworthiness, and adaptation cut across requirements engineering, testing, and software operations. Our results suggest that advancing AI-driven architecture will require interdisciplinary approaches that combine symbolic reasoning with data-driven learning and that draw from explainable AI, software analytics, and human--computer interaction. Aligning automated recommendations with human judgment and socio-technical goals is as critical in architecture as it is in the wider AI community.

\subsection{Limitations and complementary perspective on SA4AI}
At the same time, limitations of our study must be acknowledged. Our search strategy focused explicitly on the term AI, which may have caused us to miss studies that apply large language models or other emerging techniques without framing them under this label. This risk is mitigated by extending the search until August 2025 and adding manual checks in leading architecture venues, but omissions remain possible. Another boundary of scope is that we did not review work on software architecture for AI systems (SA4AI). As outlined in Section~\ref{sec:synt}, our synthesis identified three thematic clusters, but we limited our detailed analysis to the first two (AI for Architecture Design and Decision-Making, and AI for Architecture Evolution and Adaptation). The third cluster, Architecture for AI Systems, was only briefly mentioned; here we revisit it to position our work more clearly.

A complementary body of research investigates how architectures can support the design, deployment, and operation of AI-intensive applications. As summarized in Table~\ref{tab:sa4ai}, these works can be grouped into five recurring topical strands. First, several studies propose reference architectures and reusable patterns for AI systems, particularly in the domains of federated learning and foundation model agents. These aim to standardize the architectural building blocks of AI solutions, ensuring interoperability and scalability. Second, a growing line of work targets responsible, explainable, and green AI, reflecting the increasing demand for architectures that embed societal values such as transparency, accountability, and sustainability. Third, many contributions are domain-specific, addressing contexts such as smart cities, healthcare, or manufacturing, where AI places unique requirements on system integration and architectural quality attributes. Fourth, researchers have investigated architectures for AI pipelines and large-scale data processing, where challenges of scalability, modularity, and maintainability dominate. Finally, a number of studies emphasize human--AI collaboration, proposing architectures that support mixed-initiative decision making and fluid human--AI teaming. While valuable, these studies address a different, but complementary question: how AI systems themselves should be architected to ensure qualities such as scalability, responsibility, and human--AI collaboration. In contrast, our study focuses on how AI techniques can be applied to support architectural work in general software systems. The intersection of these two perspectives, AI4SA and SA4AI, offers promising opportunities for cross-fertilization. For example, techniques developed in SA4AI for explainability or human--AI teaming could inform the design of more trustworthy AI-assisted architectural capabilities, while advances in AI for SA may in turn provide automation and intelligence to the complex pipelines used in building AI systems.
\begin{table*}[htbp!]
\begin{small}
\begin{center}
\caption{Synthesis of studies on the interplay between SA and AI}
\label{tab:sa4ai}
\begin{tabular}{| p{1.6cm} | p{1cm} | p{8cm} | p{3cm} |}
 \hline
\textbf{Contribution} & \textbf{Cluster} & \textbf{Topic} & \textbf{Study} \\
   \hline\hline
   \multirow{5}{*}{SA4AI} & \multirow{5}{*}{SA4AI} 
   & Reference architectures and patterns for AI systems & \cite{P1,P32,P43,P44,P28,P49} \\
   \cline{3-4}
   & & Responsible, explainable, and green AI architectures & \cite{P3,P9,P12,P41} \\
   \cline{3-4}
   & & Domain- and application-specific AI architectures & \cite{P5,P18,P19,P22,P46} \\
   \cline{3-4}
   & & Architectures for AI pipelines and data processing & \cite{P13,P50,P23} \\
   \cline{3-4}
   & & Architectural support for human–AI collaboration and decision making & \cite{P15,P17,P45,P51} \\
   \hline
\end{tabular}
\end{center}
\end{small}
\end{table*}

Finally, although our mapping procedure was designed to ensure rigour, the alignment between practitioner challenges and AI contributions inevitably involves interpretation, and different researchers might produce slightly different categorizations. Despite this, the combination of a systematic literature review with empirically grounded practitioner challenges provides a robust basis for charting a research agenda. The articulation of AI-specific challenges clarifies where progress is most needed, while the five strategic pillars in Section~\ref{sec:roa} outline how the community can move toward a vision of AI-driven architecture that is adaptive, trustworthy, and evidence-based. Realizing this vision will require closer collaboration with industry, benchmarks and datasets that support rigorous and reproducible evaluation, and approaches that keep human expertise at the center of AI-supported architectural capabilities. Achieving this will not only address the immediate gaps identified in this study but also establish software architecture as a leading domain for the responsible adoption of artificial intelligence.

\section{Threats to validity} \label{sec:threats}
External validity refers to the extent to which our findings can be generalized beyond the primary studies included in this review. The selected studies may not fully capture all relevant literature on AI for SA. To mitigate this risk, we queried four major software engineering databases and indexing systems (\textit{IEEE Xplore}, \textit{ACM Digital Library}, \textit{Scopus}, \textit{Web of Science}) using a broad, intentionally simple search string to maximize coverage. We complemented the automatic search with manual screening of ICSA and ECSA proceedings (including companion proceedings) from 2019–2025, and applied backward and forward snowballing. Although our search string and inclusion/exclusion criteria could theoretically omit certain works, the deliberate choice to avoid enumerating specific AI techniques reduced the risk of biasing results toward particular methods. The inclusion of both academic and industrial venues and a broad publication time window further increased coverage.
Our study restricts the primary corpus to peer-reviewed publications to ensure a consistent level of methodological quality and comparability across studies. However, this choice may limit the coverage of very recent developments, particularly in a fast-evolving domain such as AI for software architecture, where influential ideas are often first disseminated through preprints (e.g., arXiv) or technical reports. To mitigate this limitation, we extended the search period to August 2025 and complemented it with manual checks of leading venues; nevertheless, we acknowledge that excluding non–peer-reviewed sources may affect the completeness of the captured landscape.

Internal validity concerns possible systematic errors in data collection or selection. To reduce this risk, we followed the guidelines for systematic secondary studies by Kitchenham et al.~\cite{kitchenham2013systematic} and by Ali and Petersen~\cite{ali2014evaluating}. The inclusion and exclusion criteria were defined before screening and applied consistently. All authors participated in key selection decisions, and disagreements were resolved by consensus. 
We refined the data extraction schema collaboratively during early stages of the extraction process. Cross-checking of coded data was performed iteratively by all authors to detect and resolve inconsistencies, and coding decisions were revisited whenever disagreements or ambiguities arose.
Internal validity is limited by the tool selection process itself that was assembled through a lightweight grey-literature analysis informed by our experience and was intended to provide a representative snapshot rather than an exhaustive catalogue, which means that some relevant tools or capabilities may have been omitted.

Construct validity addresses whether our study design accurately captures the concepts under investigation. A key threat was that an overly narrow search string might fail to retrieve relevant studies. We mitigated this by employing an inclusive string requiring minimal database-specific adaptation, supplemented by targeted manual searches of key SA venues. In the qualitative synthesis phase, grounded theory (open and axial coding) was applied inductively and iteratively to ensure categories emerged from the data rather than from preconceived notions~\cite{grounded_theory}. Maintaining explicit traceability between codes, axial categories, and their originating studies helped ensure that constructs reflected actual evidence.
 Construct validity may be affected by the fact that the characterization of tool capabilities relies primarily on publicly available vendor information, such as official websites and documentation pages, which may emphasize advertised features rather than independently validated use in practice. In addition, the Intelligence Support Levels and the mappings to architectural challenges are analytical constructs derived through the grounded and inductive reasoning process, and therefore may remain partially dependent on researcher interpretation.

Conclusion validity relates to the credibility of our interpretations and the degree to which they are supported by the data. We safeguarded this by using a transparent, well-documented analysis process, making a replication package publicly available (including search strings, selection decisions, extraction forms, and coded data). The integration of empirical SA challenges with literature findings was evidence-driven. All authors participated in mapping axial categories to challenges and in identifying AI-specific challenges, which were grounded in both data and domain expertise.
 Conclusion validity is limited by the diagnostic purpose of the tool selection as the results are intended to reveal recurring maturity patterns and capability gaps across categories of tools, rather than to support definitive comparative claims or rankings of individual vendors.

Reliability concerns the repeatability of the study. We documented each methodological step in detail, including search strategy, inclusion/exclusion criteria, coding procedures, and mapping methodology, to support replication. The replication package ensures that other researchers can reproduce our results or extend our work with minimal ambiguity.

Researcher bias is a potential risk in qualitative synthesis, particularly in interpreting how AI practices map to SA challenges. We mitigated this by involving all five authors in the open coding, axial coding, and mapping stages. These were conducted iteratively, with repeated reviews until consensus was reached. Disagreements were resolved through discussion, and all interpretations were explicitly linked to evidence from the primary studies. Considering multiple AI practices for each challenge also reduced the likelihood of overemphasising single techniques.

\section{Related Work} \label{sec:related}
AI has become a pervasive theme across software engineering research, with an increasing number of studies exploring how AI may reshape design, development, and maintenance practices.
This growing interest has resulted in a wide spectrum of contributions, but also in significant heterogeneity in quality and scope. 
To the best of our knowledge, this is the first peer-reviewed systematic literature review that consolidates evidence on AI contributions to SA, explicitly aligning these contributions with empirically derived practitioner challenges.

A large share of recent publications appear only as preprints (e.g., on arXiv), which, while often containing valuable early insights, have not undergone peer review and therefore lack the reliability and permanence required for a systematic account.\footnote{For instance, Esposito et al., \url{https://papers.ssrn.com/sol3/papers.cfm?abstract_id=5196419}}
In contrast, our related work discussion deliberately restricts itself to peer-reviewed studies to ensure methodological rigour and replicability

Several studies have investigated the intersection of AI and architecture, but typically with  narrower lens than adopted here. For example, Rittelmeyer and Sandkuhl conducted a structured literature review on the effects of AI applications on enterprise architectures~\cite{rittelmeyer2021effects}. Their results highlight that AI adoption has consequences ranging from changes in data formats to full process restructuring, yet the research landscape remains fragmented and lacking concrete insights into architectural practices. 
Similarly, Jahic et al. reported on a state-of-practice survey in the embedded systems domain, showing that while companies perceive AI adoption as highly disruptive, the architectural techniques available to anticipate and manage such disruption are still immature~\cite{jahic2020state}.
These studies provide valuable evidence in their respective contexts, but do not offer a holistic view of AI’s role in software architecture as a whole.

Beyond these works, the majority of reviews and secondary studies focus either on individual AI techniques (e.g., generative AI for code or design tasks) or on narrow architectural concerns (e.g., documentation, adaptation, or performance optimization). 
An example of this is the study by Saucedo and Rodríguez, which systematically mapped AI-supported approaches for migrating monolithic systems to microservices~\cite{saucedo2024migration}. Their review of 22 studies found that clustering was the most common AI technique (used in 63\% of cases) for decomposing monoliths, with source code (36.4\%) being the prevailing input type. While valuable, such work remains narrow in scope and does not generalize to other architectural concerns.

A further strand of secondary studies situates AI within the broader software engineering landscape, where architecture is considered only peripherally. 
For instance, a recent systematic mapping review examined how generative AI tools such as GitHub Copilot and ChatGPT impact software development processes~\cite{santos2024impacts}. The study found that the strongest evidence of adoption and benefits lies in development and testing activities, with more limited insights into requirements specification and architecture. While highlighting productivity gains over traditional practices like pair programming or manual testing, the review also notes unresolved issues around reliability and applicability. Similar studies exist for testing, requirements engineering, and project management, reinforcing that while AI’s influence across SE is widely recognized, architecture is rarely the focal point of analysis.

 Historically, influential roadmap papers have articulated broad, discipline-level research agendas for software architecture~\cite{garlan2014software, kramer2007self}. These contributions provide expert-driven perspectives spanning historical evolution, socio-technical considerations, and long-term technological drivers such as network-centric and pervasive computing. Importantly, they also fostered a substantial body of work on self-adaptive and autonomic architectures, which explored architectural mechanisms for managing uncertainty and dynamically adapting QAs at runtime.

These self-managed approaches typically rely on explicit architectural models and feedback control loops (e.g., Sense–Plan–Act or MAPE-K) to enable self-healing, self-tuning, and self-optimizing behaviours. This tradition represents a foundational milestone in the evolution of intelligent software architectures and has significantly advanced runtime adaptation under uncertainty.

Our work adopts a deliberately different scope and methodological foundation. Rather than advancing a general research agenda for software architecture as a whole, we focus specifically on the explicit role of contemporary AI techniques in architectural practice. In light of the rapid and transformative growth of data-driven and generative AI, we examine how these technologies act as catalysts for architectural change across the lifecycle, beyond runtime control mechanisms. Methodologically, our roadmap is grounded in a systematic literature review of 51 primary studies and an explicit alignment with empirically derived practitioner challenges.

In this sense, our roadmap should be understood as a maturation study that evaluates how modern AI techniques extend—and where they still fall short of—the long-standing vision of intelligent and self-managing architectures, particularly with respect to lifecycle-wide reasoning, continuous traceability, and long-term architectural evolution.

All together, these fragmented perspectives show that prior work either emphasizes specific domains (e.g., enterprise or embedded systems), individual architectural tasks (e.g., migration), or broader software engineering contexts where architecture is secondary. Our study addresses this gap by systematically consolidating peer-reviewed evidence across 51 primary studies and aligning these findings with 17 practitioner-reported challenges. This dual focus allows us to provide the first comprehensive and empirically grounded synthesis of how AI contributes to SA, where its limitations lie, and how research can move toward a vision of AI-driven software architecture.
\section{Conclusion and Future Work} \label{sec:conclusion}
AI is beginning to transform SA, but its current contributions remain fragmented, reactive, and often detached from real-world practice. 
In this paper, we presented the first peer-reviewed systematic literature review consolidating 51 primary studies on AI for SA, and systematically aligned their findings with 17 empirically derived challenges reported by practitioners. This mapping allowed us to distil six AI-specific challenges (AICHs) that represent the frontier of AI-driven SA.

Our analysis shows that while promising AI techniques exist for tasks such as design support, documentation recovery, performance estimation, and quality prediction, they are typically limited to static, one-off recommendations, narrow domains, or controlled settings. In contrast, practitioners face evolving requirements, debt-laden systems, and organizational contexts that demand continuous, explainable, and trustworthy support. The AICHs crystallize this mismatch by identifying where AI techniques must evolve—from static outputs to adaptive learning, from isolated analyses to system-wide reasoning, and from abstract models to contextualized, human-centred collaboration.

 In this sense, our roadmap should be understood as a maturation study that evaluates how contemporary AI techniques extend, and where they still fall short of, the long-standing vision of intelligent and self-managing architectures.

Building on these findings, we proposed a forward-looking vision for AI-driven SA structured around five strategic pillars: continuous and adaptive architecting, living architectural knowledge, context-aware and human-centred AI, evidence-based quantitative architecting, and benchmarks and industrial adoption. Together, these pillars chart a research roadmap that connects today’s prototypes with tomorrow’s robust and trustworthy AI-enabled practices.

For researchers, our work provides a rigorous foundation for identifying gaps and opportunities, highlighting where new methods, datasets, and evaluation protocols are most urgently needed. For practitioners, it clarifies both the potential and current limitations of AI techniques, guiding realistic expectations of their utility in practice. Future research will need to bridge this divide through stronger industrial collaborations, continuous validation in evolving ecosystems, and hybrid approaches that integrate AI capabilities with human expertise.

\begin{acks}
This work is supported by the Swedish Agency for Innovation Systems through
the projects "iSecure: Developing Predictable and Secure IoT for Autonomous Systems" (2023-01899), and "Advancing AI-Driven Software Architecture: A Collaborative Exchange between MDU and SMU" (2024-02068), by the Key Digital Technologies Joint Undertaking through the project "MATISSE: Model-based engineering of digital twins for early verification and validation of industrial systems" (101140216), and by the Clean Energy Transition Partnership through the project ``FLEXI: Human-centered AI and digital twin powered energy system integration for flexibility markets" (101069750).
This research / project is also supported by the National Research Foundation, under its Investigatorship Grant (NRF-NRFI08-2022-0002). Any opinions, findings and conclusions or recommendations expressed in this material are those of the author(s) and do not reflect the views of National Research Foundation, Singapore.
\end{acks}

\clearpage
\renewcommand\refname{Primary studies} \label{sel}

\renewcommand\refname{References}

\bibliographystyle{ACM-Reference-Format}
\bibliography{software}

\appendix


\section{Online Resources} \label{rep}
To facilitate independent replication and verification, we provide a complete replication package 
\url{https://doi.org/10.5281/zenodo.20798468}
containing search and selection data, and the list of primary studies.

\end{document}